\documentclass[twocolumn]{aastex62}

\usepackage[shortcuts]{extdash}

\def\apjl{ApJL}

\def\aap{A\&A}

\def\apjs{ApJS}

\def\arcsec{$^{\prime\prime}$}

\received{}
\revised{}
\accepted{}
\submitjournal{} 

\shorttitle{ULXs in NGC 6946}
\shortauthors{Earnshaw et al.}

\begin{document}

\title{A broadband look at the old and new ULXs of NGC 6946}

\correspondingauthor{Hannah P. Earnshaw}
\email{hpearn@caltech.edu}

\author[0000-0001-5857-5622]{Hannah P. Earnshaw}
\affil{Cahill Center for Astronomy and Astrophysics, California Institute of Technology, Pasadena, CA 91125, USA}

\author{Brian W. Grefenstette}
\affil{Cahill Center for Astronomy and Astrophysics, California Institute of Technology, Pasadena, CA 91125, USA}

\author{Murray Brightman}
\affil{Cahill Center for Astronomy and Astrophysics, California Institute of Technology, Pasadena, CA 91125, USA}

\author{Dominic J. Walton}
\affil{Institute of Astronomy, Madingley Road, Cambridge CB3 0HA, UK}

\author{Didier Barret}
\affil{IRAP, Universit\'{e} de Toulouse, CNRS, UPS, CNES, Toulouse, France}

\author{Felix F\"{u}rst}
\affil{European Space Astronomy Centre (ESAC), Science Operations Department, 28692, Villanueva de la Cañada, Madrid, Spain}

\author{Fiona A. Harrison}
\affil{Cahill Center for Astronomy and Astrophysics, California Institute of Technology, Pasadena, CA 91125, USA}

\author{Marianne Heida}
\affil{Cahill Center for Astronomy and Astrophysics, California Institute of Technology, Pasadena, CA 91125, USA}

\author{Sean N. Pike}
\affil{Cahill Center for Astronomy and Astrophysics, California Institute of Technology, Pasadena, CA 91125, USA}

\author{Daniel Stern}
\affil{Jet Propulsion Laboratory, California Institute of Technology, 4800 Oak Grove Drive, Pasadena, CA 91109, USA}

\author{Natalie A. Webb}
\affil{IRAP, Universit\'{e} de Toulouse, CNRS, UPS, CNES, Toulouse, France}



\begin{abstract}

Two recent observations of the nearby galaxy NGC~6946 with {\it NuSTAR}, one simultaneous with an {\it XMM-Newton} observation, provide an opportunity to examine its population of bright accreting sources from a broadband perspective. We study the three known ultraluminous X-ray sources (ULXs) in the galaxy, and find that ULX\=/1 and ULX\=/2 have very steep power-law spectra with $\Gamma=3.6^{+0.4}_{-0.3}$ in both cases. Their properties are consistent with being super-Eddington accreting sources with the majority of their hard emission obscured and down-scattered. ULX\=/3 (NGC~6946~X\=/1) is significantly detected by both {\it XMM-Newton} and {\it NuSTAR} at $L_{\rm X}=(6.5\pm0.1)\times10^{39}$\,erg\,s$^{-1}$, and has a power-law spectrum with $\Gamma=2.51\pm0.05$. We are unable to identify a high-energy break in its spectrum like that found in other ULXs, but the soft spectrum likely hinders our ability to detect one. We also characterise the new source, ULX\=/4, which is only detected in the joint {\it XMM-Newton} and {\it NuSTAR} observation, at $L_{\rm X}=(2.27\pm0.07)\times10^{39}$\,erg\,s$^{-1}$, and is absent in a {\it Chandra} observation ten days later. It has a very hard cut-off power-law spectrum with $\Gamma=0.7\pm0.1$ and $E_{\rm cut}=11^{+9}_{-4}$\,keV. We do not detect pulsations from ULX\=/4, but its transient nature can be explained either as a neutron star ULX briefly leaving the propeller regime or as a micro-tidal disruption event induced by a stellar-mass compact object.

\end{abstract}


\section{Introduction} \label{sec:intro}

The study of ultraluminous X-ray sources (ULXs) offers insight into some of the most extreme accretion processes in the Universe (for a recent review, see \citealt{kaaret17}). This is indicated by their high luminosities ($L_{\rm X} > 10^{39}$\,erg\,s$^{-1}$) which imply either the presence of an intermediate-mass black hole (IMBH; $10^2<M_{\rm BH}<10^5$\,M$_{\odot}$; \citealt{colbert99}), or apparent super-Eddington accretion onto a stellar-mass compact object (e.g. \citealt{sutton13b,bachetti14}). In recent years, broadband X-ray observations using {\it XMM-Newton} and {\it NuSTAR} have shown the spectral and timing properties of ULXs to be distinct from those of sub-Eddington accretion states, with characteristic broadened-disc or two-component shapes in the {\it XMM-Newton} band, a spectral turnover at $\gtrsim5$\,keV and, in the case of especially good {\it NuSTAR} data above $\sim20$\,keV, an additional steep power-law excess at the highest energies (e.g. \citealt{stobbart06,gladstone09,bachetti13,mukherjee15,walton14,walton15b}). These properties are consistent with a supercritical accretion regime onto a stellar-mass compact object, in which a cooler component is emitted from an outer supercritical accretion disc launching an outflowing wind, and a hotter component is emitted from the inner accretion flow (e.g. \citealt{middleton15a}). Which component is dominant in the spectrum depends on the inclination and/or mass accretion rate in this model, and at very high inclinations/accretion rates the high-energy emission may be completely obscured and reprocessed, leaving an ultraluminous supersoft source with a spectrum entirely dominated by a soft thermal component (e.g. \citealt{distefano04,urquhart16,earnshaw17}). Conversely, ULXs with spectra dominated by hard emission are more likely to be sources observed at low inclinations in this picture, and make particularly good targets for {\it NuSTAR} given its sensitivity to energies $>10$\,keV. 

It was initially assumed that, due to their extreme luminosities, these stellar-mass objects would none-the-less be as massive as reasonably possible, and therefore be black holes (BHs) by default. However, the detection of pulsations and cyclotron absorption features from a number of ULXs has shown at least some fraction of ULXs to be neutron stars (NSs) instead, with luminosities 100--1000 times their isotropic Eddington limit (e.g. \citealt{bachetti14,fuerst16,israel17a,israel17b,brightman18,carpano18}; Rodr\'{i}guez Castillo et al. subm.; Sathyaprakash et al. subm.). Those ULXs confirmed to be NSs share the property of a hard spectrum which can be modeled as being dominated by a cut-off power-law emission component originating from the accretion column and providing the pulsed portion of the spectrum \citep{brightman16,walton18a}. While pulsations have not been detected from all ULXs, all of the highest-quality spectra of super-Eddington-type ULXs are potentially consistent with various different NS accretion models \citep{koliopanos17,pintore17,walton18b}.

\begin{deluxetable}{lccl}
	\tablecaption{Four ULXs in NGC~6946. \label{tab:srcs}}
	\tablecolumns{5}
	\tablenum{1}
	\tablewidth{0pt}
	\tablehead{
		\colhead{Name\tablenotemark{a}} & \colhead{Position (J2000)} & \colhead{{\it NuSTAR}?\tablenotemark{b}} & \colhead{Other names}
	}
\startdata
ULX-1 & 20:35:00.3 +60:09:07 & N & - \\
ULX-2 & 20:34:36.5 +60:09:30 & N & - \\
ULX-3 & 20:35:00.7 +60:11:31 & Y & X-1$^1$, MF16$^2$ \\
ULX-4 & 20:34:56.9 +60:08:13 & Y & - \\
\enddata
\tablenotetext{a}{The source name used in this paper, as defined in \citet{liu05} for the first three ULXs.}
\tablenotetext{b}{Detected by {\it NuSTAR}.}
\tablecomments{$^1$\citet{fabbiano87}, most commonly used name, $^2$ Associated nebula \citep{matonick97}}
\vspace{-8mm}
\end{deluxetable}

Most ULXs are persistently bright sources, some of which exhibit high levels of long-term, inter-observation variability (e.g. \citealt{fridriksson08,sutton12,grise13}). This can sometimes be sufficient to cross the $L_{\rm X}\geq10^{39}$\,erg\,s$^{-1}$ boundary which is the empirical definition of a ULX, so that a source can appear as a ULX in some observations and not others (e.g. \citealt{lin13,earnshaw17}). More dramatically, a small handful of ULXs will sometimes drop in flux by orders of magnitude or so much as to become undetectable (e.g. \citealt{walton15a,earnshaw18}). This latter scenario is observed in several of the ULX pulsars discovered to date, giving them an approximately bimodal light curve \citep{tsygankov16,fuerst16,israel17b}. This can be explained by the `propeller effect', in which accretion is stopped during periods when the magnetospheric radius of the NS exceeds the corotation radius of the accretion disc \citep{illarionov75,stella86}. ULXs have also been known to appear (e.g. \citealt{soria12,pintore18}) or disappear (often the case for classical outbursts that briefly exceed $10^{39}$\,erg\,s$^{-1}$ before returning to a quiescent state; e.g. \citealt{middleton13}).

NGC~6946 is a nearby spiral galaxy, located at a distance of 7.72\,Mpc \citep{anand18} and containing three previously detected ULXs within its spiral arms \citep{liu05}, including the well-studied soft and variable NGC~6946~X\=/1 \citep{roberts03,holt03,fridriksson08,rao10,berghea12,berghea13}, referred to as ULX\=/3 in this paper, as in \citet{liu05}, to prevent confusion with ULX\=/1. ULX\=/3 is also an ultraluminous ultraviolet source \citep{kaaret10}, is associated with the optical nebula MF16 \citep{matonick97,abolmasov08}, and shows evidence of emission lines that may be consistent with collisional heating due to an outflowing wind \citep{pinto16}. It is often referred to in literature as the only ULX of NGC~6946, as the others are less frequently observed at ULX luminosities. However, NGC~6946 also contains the soft sources NGC~6946~ULX\=/1 \citep{devi08,earnshaw17} and ULX\=/2 \citep{liu05} that have previously been observed as ULXs. 

In this paper we report on the known ULXs in NGC~6946, including ULX\=/3 which we can study with broadband X-ray data for the first time. We also report on the appearance (and later disappearance) of a new source that we call ULX\=/4, and perform multi-wavelength analysis of the region to attempt to discern its nature.

\section{Data Reduction and Analysis} \label{sec:data}

In 2017, NGC~6946 was observed twice by {\it NuSTAR} as a target of opportunity (ToO) observation in order to study the type II-P supernova SN2017eaw in the north of the galaxy (Grefenstette et al. in prep). The two observations were taken 11 days apart, with the second observation being simultaneous with an observation by {\it XMM-Newton}. During these observations, all three previously identified ULXs were detected by {\it XMM-Newton}, and ULX\=/3 was detected by {\it NuSTAR}. In the second, simultaneous epoch a fourth ULX was detected by both {\it XMM-Newton} and {\it NuSTAR}, having newly appeared in the 11 days between the observations. We present a list of the ULXs in NGC~6946 in Table~\ref{tab:srcs}. 

\begin{figure*}
	\begin{center}
	\includegraphics[height=75mm,trim={0 0 4mm 0},clip]{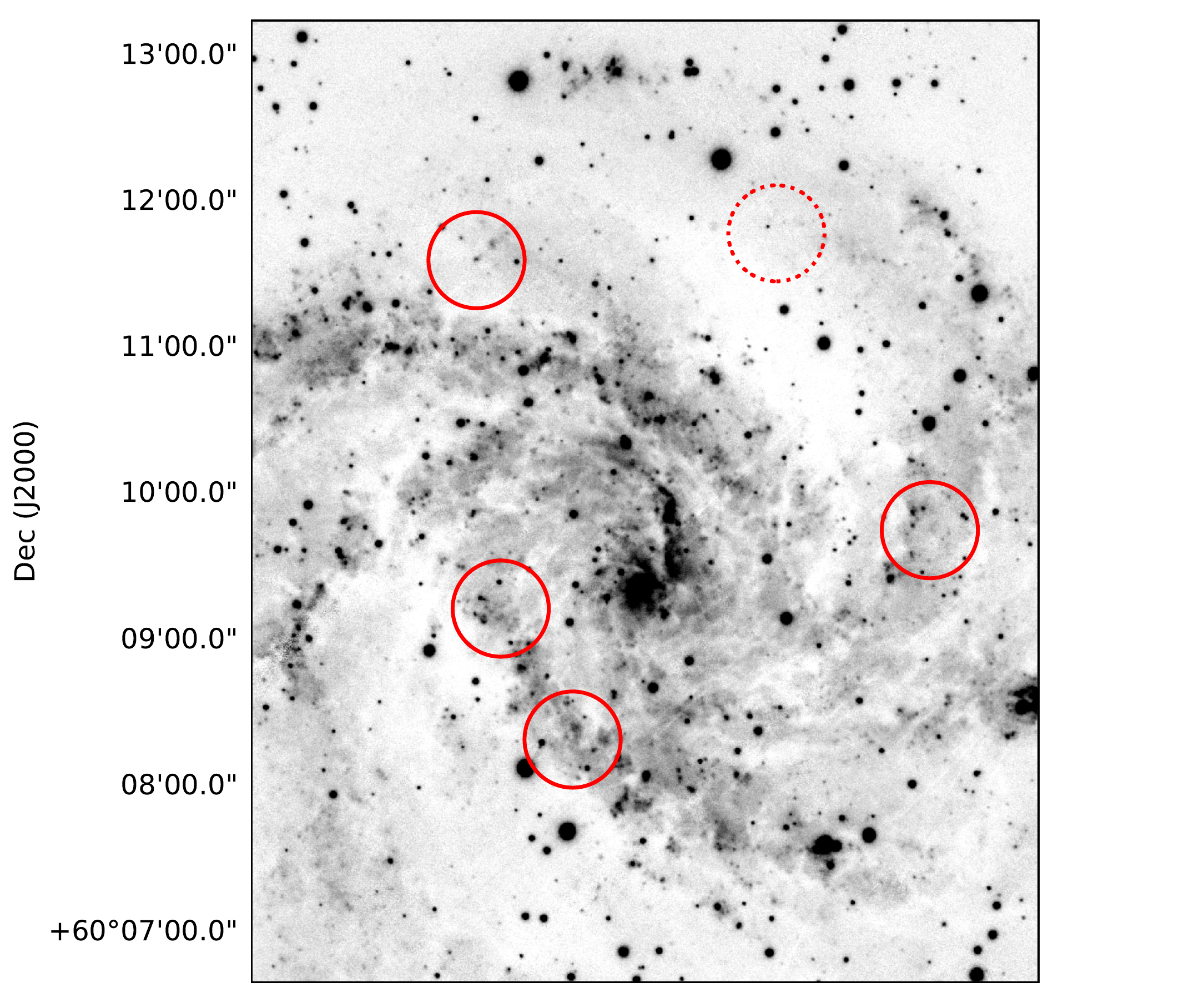}
	\hspace{-1cm}
	\includegraphics[height=75mm,trim={28mm 0 0 0},clip]{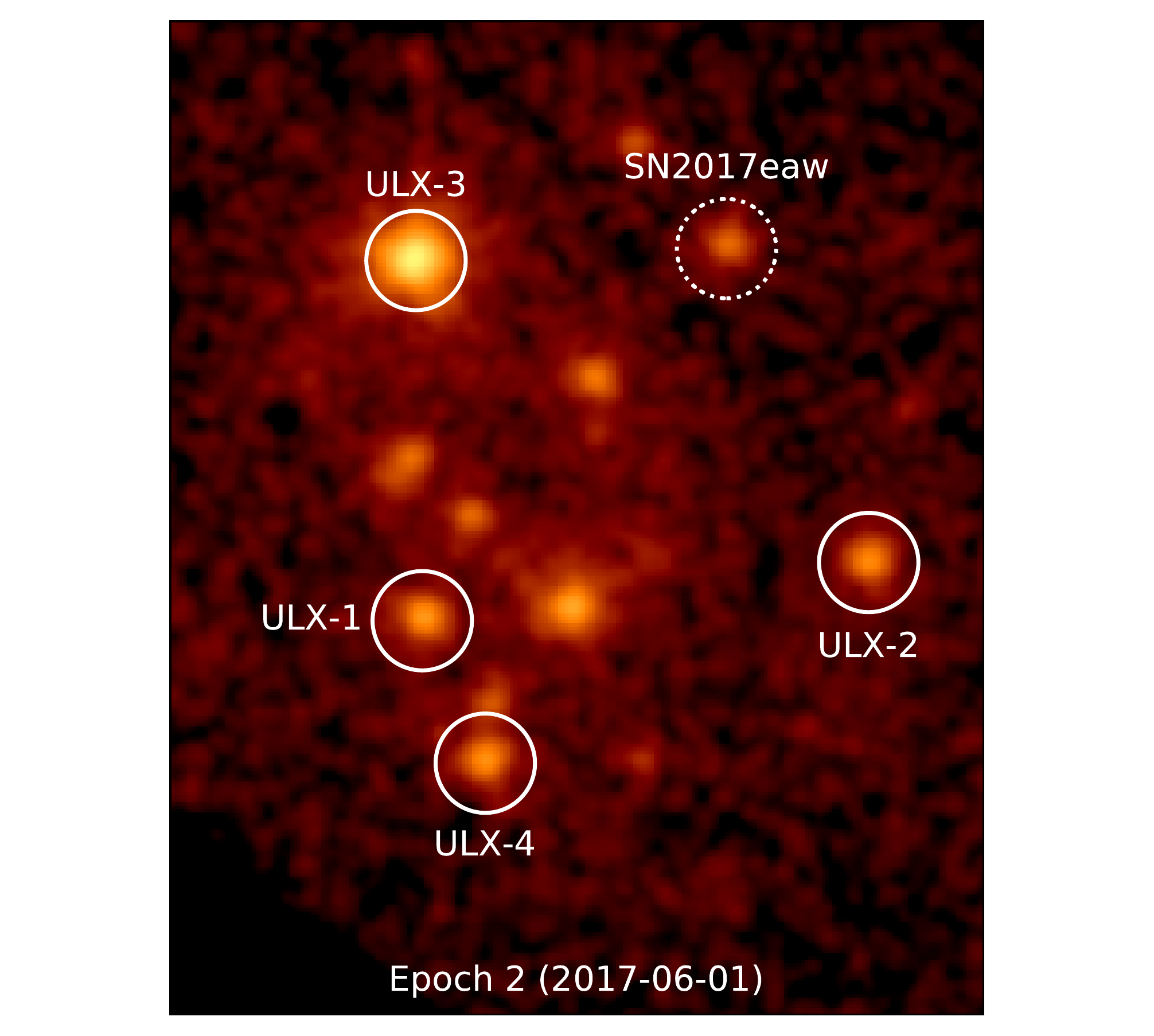}
	\includegraphics[height=80mm,trim={0 0 4mm 0},clip]{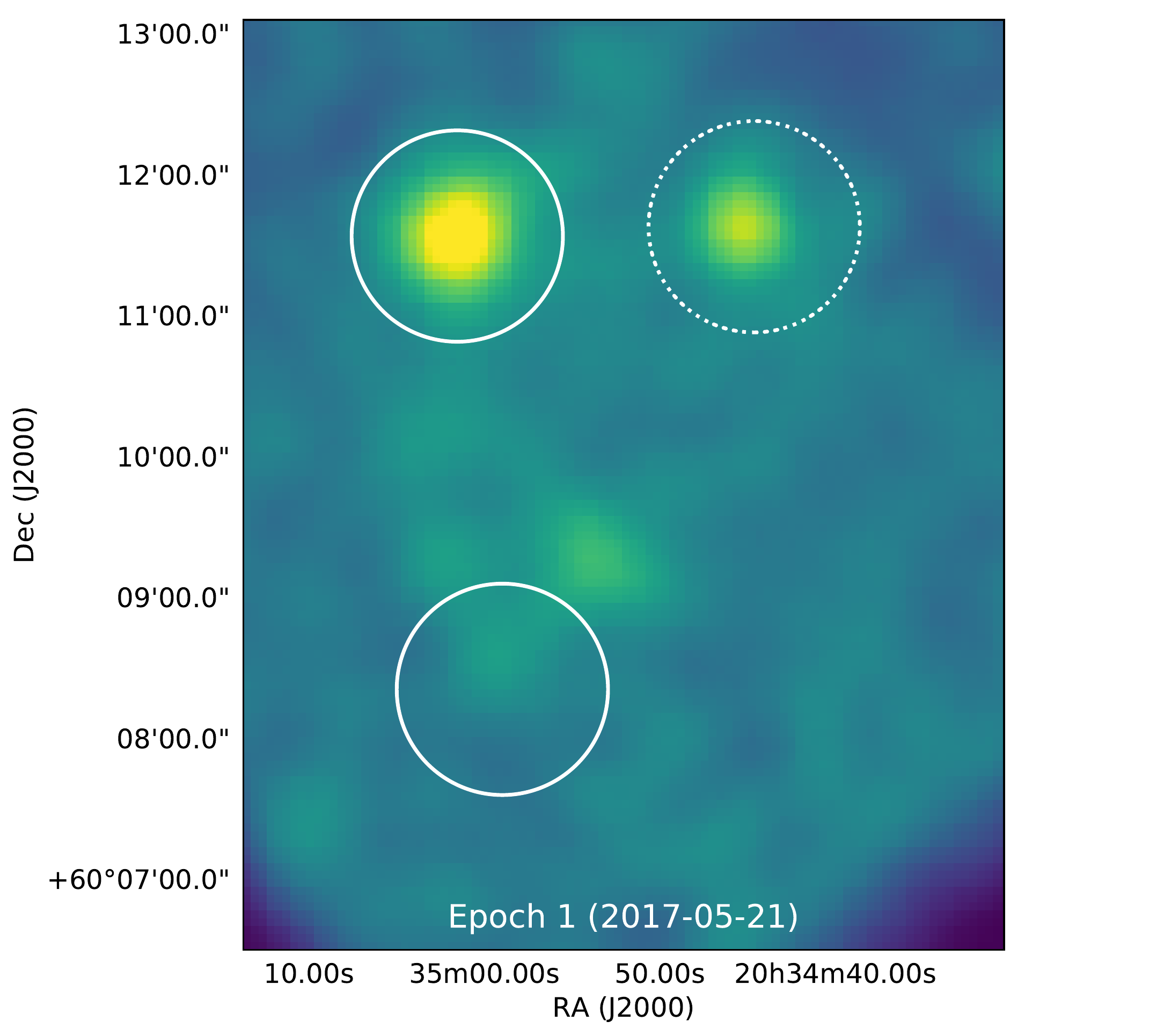}
	\hspace{-1cm}
	\includegraphics[height=80mm,trim={28mm 0 0 0},clip]{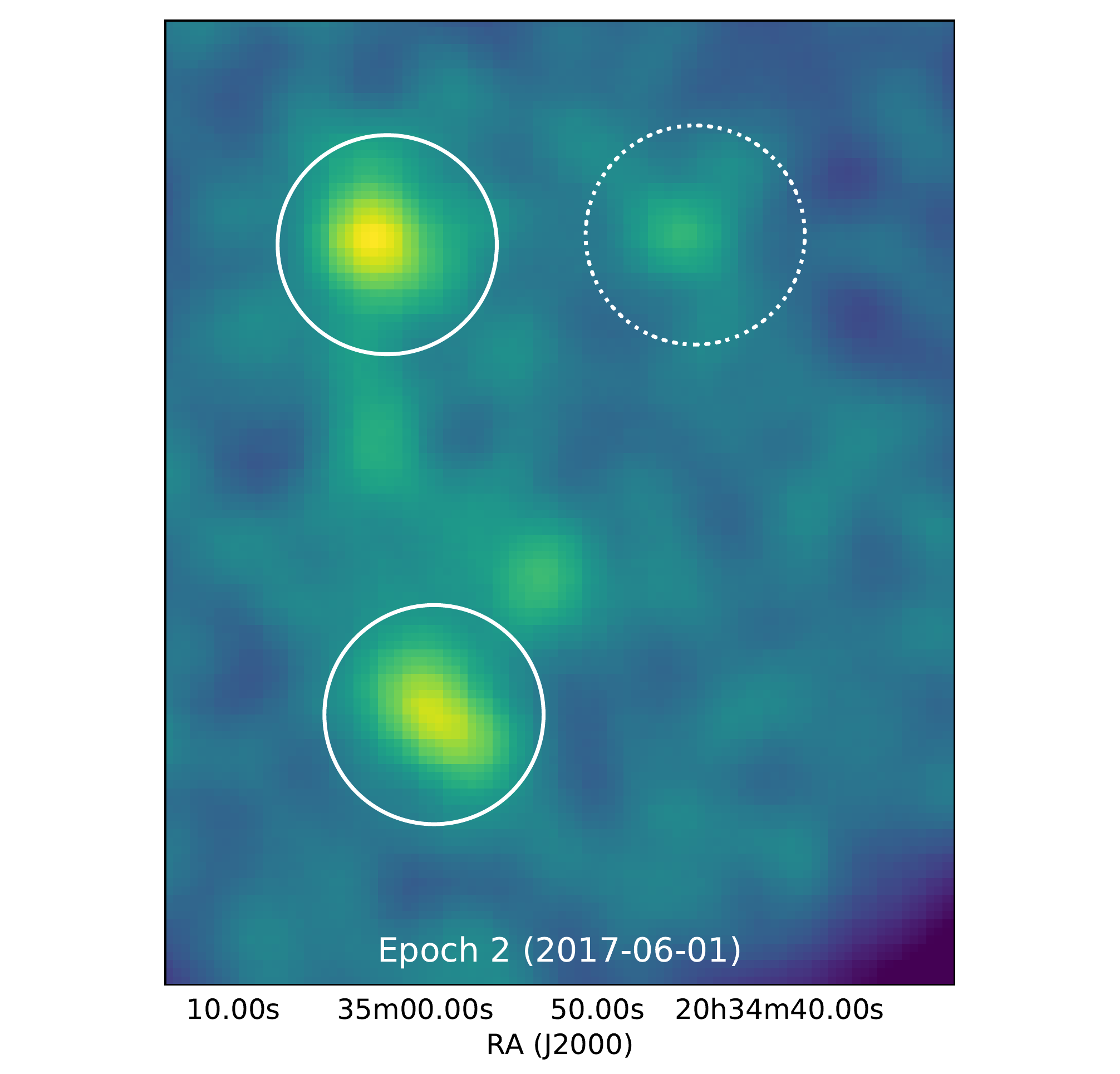}
	\end{center}
	\caption{Optical and X-ray observations of the galaxy NGC~6946. {\it Top left,} optical $g$-band {\it Pan-STARRS} image with the locations of the ULXs and SN2017eaw marked with 20\arcsec\ red circles (SN2017eaw's circle is dashed). {\it Top right,} soft X-ray image of {\it XMM-Newton} observation 0794581201 with the locations of the ULXs and SN2017eaw labeled marked with 20\arcsec\ white circles. {\it Bottom left,} hard X-ray image of {\it NuSTAR} observation 90302004002, with the locations of SN2017eaw, ULX\=/3, and the ULX\=/4 non-detection marked with 45\arcsec\ white circles. {\it Bottom right,} hard X-ray image of {\it NuSTAR} observation 90302004004, with the locations of SN2017eaw, ULX\=/3, and ULX\=/4 marked with 45\arcsec\ white circles. \label{fig:obs}}
	\vspace{4mm}
\end{figure*}

In our analysis of the ULXs in NGC~6946, we primarily use the two {\it NuSTAR} observations and the associated {\it XMM-Newton} observation simultaneous with the second {\it NuSTAR} observation, all taken in 2017 as part of the same ToO campaign. In examining the history of ULX\=/4, we also make use of all archival {\it Chandra} observations, archival {\it XMM-Newton} observations, and all {\it Swift} observations taken in 2017. We present a list of the X-ray observations used in this study in Table~\ref{tab:obs}, with the 2017 ToO observations marked in bold. We show images of the 2017 {\it XMM-Newton} and {\it NuSTAR} observations in Fig.~\ref{fig:obs}, along with an optical image of NGC~6946.

\begin{deluxetable}{@{~}c@{~~}c@{~~~}c@{~~}c@{~}}
	\tablecaption{2017 and archival X-ray observations of NGC~6946. \label{tab:obs}}
	\tablecolumns{4}
	\tablenum{2}
	\tablewidth{0pt}
	\tablehead{
		\colhead{Obs. ID} & \colhead{Mission} & \colhead{Obs. Date} & \colhead{Exposure\tablenotemark{a} (ks)}
	}
	\startdata
	\multicolumn{4}{c}{\textbf{\textit{2017 ToO observations}}} \\
	\bf 90302004002 & \textbf{\textit{NuSTAR}} & \bf 2017-05-21 & \bf 66.7 \\
	\bf 90302004004 & \textbf{\textit{NuSTAR}} & \bf 2017-06-01 & \bf 47.7 \\
	\bf 0794581201 & \textbf{\textit{XMM-Newton}} & \bf 2017-06-01 & \bf 44.0/45.2/39.2 \\ 
	\hline
	0200670101 & {\it XMM-Newton} & 2004-06-09 & 4.0/4.1/0.8 \\ 
	0200670301 & {\it XMM-Newton} & 2004-06-13 & 12.4/12.4/8.3 \\
	0200670401 & {\it XMM-Newton} & 2004-06-25 & 8.6/9.4/0.8 \\
	0401360201 & {\it XMM-Newton} & 2006-06-02 & 4.6/4.4/0.1 \\
	0401360301 & {\it XMM-Newton} & 2006-06-18 & 6.0/6.0/1.7 \\
	0500730101 & {\it XMM-Newton} & 2007-11-08 & 27.9/28.3/20.2 \\
	0500730201 & {\it XMM-Newton} & 2007-11-02 & 32.4/32.6/29.8 \\
	0691570101 & {\it XMM-Newton} & 2012-10-21 & 110.7/112.2/98.2 \\ 
	1043 & {\it Chandra} & 2001-09-07 & 58.3 \\
	4404 & {\it Chandra} & 2002-11-25 & 30.0 \\
	4631 & {\it Chandra} & 2004-10-22 & 29.7 \\
	4632 & {\it Chandra} & 2004-11-06 & 28.0 \\
	4633 & {\it Chandra} & 2004-12-03 & 26.6 \\
	13435 & {\it Chandra} & 2012-05-21 & 20.4 \\
	17878 & {\it Chandra} & 2016-09-28 & 40.0 \\
	19887 & {\it Chandra} & 2016-09-28 & 18.6 \\
	19040 & {\it Chandra} & 2017-06-11 & 9.8 \\
	00010130001 & {\it Swift-XRT} & 2017-05-14 & 2.0 \\
	00010130003 & {\it Swift-XRT} & 2017-05-15 & 1.8 \\
	00010130004 & {\it Swift-XRT} & 2017-05-15 & 3.8 \\
	00010130005 & {\it Swift-XRT} & 2017-05-16 & 1.7 \\
	00010130006 & {\it Swift-XRT} & 2017-05-17 & 2.6 \\
	00010130007 & {\it Swift-XRT} & 2017-05-18 & 1.5 \\
	00010130008 & {\it Swift-XRT} & 2017-05-22 & 4.0 \\
	00010130010 & {\it Swift-XRT} & 2017-05-24 & 1.7 \\
	00010130011 & {\it Swift-XRT} & 2017-05-26 & 1.8 \\
	00010130012 & {\it Swift-XRT} & 2017-05-29 & 0.7 \\
	00010130013 & {\it Swift-XRT} & 2017-05-29 & 1.3 \\
	00010130014 & {\it Swift-XRT} & 2017-05-31 & 1.5 \\
	00010130015 & {\it Swift-XRT} & 2017-06-02 & 1.5 \\
	00010130016 & {\it Swift-XRT} & 2017-06-04 & 1.5 \\
	00010130017 & {\it Swift-XRT} & 2017-06-06 & 1.6 \\
	00010130018 & {\it Swift-XRT} & 2017-06-08 & 1.4 \\
	00010130019 & {\it Swift-XRT} & 2017-06-10 & 1.5 \\
	00010130020 & {\it Swift-XRT} & 2017-06-12 & 0.5 \\
	00010130021 & {\it Swift-XRT} & 2017-06-14 & 1.1 \\
	00010130022 & {\it Swift-XRT} & 2017-06-16 & 1.5 \\
	00010130023 & {\it Swift-XRT} & 2017-07-15 & 3.0 \\
	00010130024 & {\it Swift-XRT} & 2017-07-28 & 3.0 \\
	00010130025 & {\it Swift-XRT} & 2017-08-11 & 1.0 \\
	00010130026 & {\it Swift-XRT} & 2017-08-16 & 1.3 \\
	00010130027 & {\it Swift-XRT} & 2017-08-25 & 3.2 \\
	00010130028 & {\it Swift-XRT} & 2017-09-08 & 1.9 \\
	00010130029 & {\it Swift-XRT} & 2017-09-13 & 0.9 \\
	\enddata
	\vspace{-2mm}
	\tablenotetext{a}{Given as EPIC-MOS1/MOS2/pn for {\it XMM-Newton}, after removal of periods of background flaring.}
\end{deluxetable}

\subsection{NuSTAR}

We reprocessed the {\it NuSTAR} observations using the NuSTAR Data Analysis Software (NuSTARDAS; v1.7.1) routine {\tt nupipeline}. We extracted source and background spectra and light curves from the {\it NuSTAR} science data with the NuSTARDAS task {\tt nuproducts}, using circular source extraction regions with radius 45\arcsec\, and background regions of radius 60\arcsec\ on the same chip. Spectra were grouped into 20 counts per bin to allow for $\chi^2$ statistics to be used in spectral model fitting. A $3\sigma$ upper limit for the initial non-detection of ULX\=/4 was found using the FTOOLS task {\tt sosta}, using a 45\arcsec\ radius source region and a 60\arcsec\ radius background region.

\subsection{XMM-Newton}

The {\it XMM-Newton} data were reduced using v16.1.0 of the {\it XMM-Newton} Science Analysis System (SAS) software and up-to-date CALDB as of June 2018, producing calibrated event files with {\tt epproc} and {\tt emproc} and removing periods of background flaring -- several observations were heavily affected by background flaring so were removed from this analysis. We extracted data products using 20\arcsec\ radius circular source regions and 40\arcsec\ radius circular background regions located on the same chip at a similar distance from the readout node. We selected {\tt FLAG==0 \&\& PATTERN$<$4} events for the EPIC-pn camera, and {\tt PATTERN$<$12} for the EPIC-MOS cameras. In all cases, spectra were grouped into 20 counts per bin as for {\it NuSTAR}. Redistribution matrices and auxiliary response files were generated with the tasks {\tt rmfgen} and {\tt arfgen} respectively. $3\sigma$ flux upper limits for non-detections were determined using the SAS task {\tt eregionanalyse}, again using 20\arcsec\ radius circular source regions and 40\arcsec\ radius circular background regions.

\begin{deluxetable*}{lcccccccc}
	\vspace{-1mm}
	\tablecaption{The ULX\=/1 and ULX\=/2 spectral fitting results for {\it XMM-Newton} observation 0794581201. \label{tab:ulx12}}
	\tablecolumns{9}
	\tablenum{3}
	\tablewidth{0pt}
	\tablehead{
		\colhead{Source} & \colhead{$N_{\rm H}$\tablenotemark{a}} & \colhead{$\Gamma$/$T_{\rm in}$} & \colhead{$p$\tablenotemark{b}} & \colhead{$T_{\rm in}$} & \colhead{$E_{\rm line}$} & \colhead{$\sigma_{\rm line}$} & \colhead{$\chi^2/{\rm dof}$} & \colhead{$F_{0.3-10 \rm keV}$\tablenotemark{c}} \\
		\colhead{} & \colhead{($10^{21}$\,cm$^{-2}$)} & \colhead{(-/keV)} & \colhead{} &  \colhead{(keV)} & \colhead{(keV)} & \colhead{(keV)} & \colhead{} & \colhead{($10^{-14}$\,erg\,s$^{-1}$)}
	}
	\startdata
	ULX-1 &  {\tt tbabs*tbabs*} & {\tt (powerlaw} & & & \multicolumn{2}{c}{\tt + gauss)} & & \\
	 & $4.2^{+0.7}_{-0.6}$ & $4.5\pm0.3$ & - & - & - & - & 131.9/90 & - \\
     & $2.3^{+0.8}_{-0.7}$ & $3.6^{+0.4}_{-0.3}$ & - & - & $0.90^{+0.04}_{-0.07}$ & $0.13^{+0.05}_{-0.04}$ & 84.5/87 & $9.8\pm0.4$ \\
     &  {\tt tbabs*tbabs*} & {\tt (diskbb} &  & & \multicolumn{2}{c}{\tt + gauss)} & & \\
	 & $0.6\pm0.4$ & $0.31\pm0.03$ & - & - & - & - & 179.7/90 & - \\
     & $<8.2$ & $0.6\pm0.1$ & - & - & $0.7\pm0.1$ & $0.29^{+0.06}_{-0.05}$ & 99.1/87 & - \\
     \hline
    ULX-2 &  {\tt tbabs*tbabs*} & {\tt powerlaw} & & & &  & & \\
	 & $2.9^{+1.1}_{-1.0}$ & $3.6^{+0.4}_{-0.3}$ & - & - & - & - & 44.9/42 & $7.8\pm0.5$ \\
	 &  {\tt tbabs*tbabs*} & {\tt (diskbb} & & {\tt + diskbb)} & & & & \\
	 & $<0.3$ & $0.43\pm0.04$ & - & - & - & - & 62.6/42 & - \\
	 & $3.3^{+0.3}_{-0.2}$ & $0.17^{+0.06}_{-0.04}$ & - & $0.7^{+0.2}_{-0.1}$ & - & - & 38.1/40 & - \\ 
	 &  {\tt tbabs*tbabs*} & \multicolumn{2}{c}{\tt diskpbb} & & & & & \\
	 & $<0.8$ & $0.7\pm0.1$ & $<0.56$ & - & - & - & 52.6/41 & - \\
	\enddata
	\tablenotetext{a}{The column density for the absorption component allowed to vary, with the first component frozen to the Galactic value of $N_{\rm H} = 1.84\times10^{21}$\,cm$^{-2}$.}
	\tablenotetext{b}{The radial dependence of disc temperature, with a hard lower limit of 0.5.}
	\tablenotetext{c}{The absorbed flux in the 0.3--10\,keV band.}
	\vspace{-1mm}
\end{deluxetable*}

\vspace{10mm}
\subsection{Chandra}

Previous {\it Chandra} observations of NGC~6946 were reduced using task {\tt chandra\_repro} of v4.7.7 of the {\it Chandra} Interactive Analysis of Observations (CIAO) software. The {\tt srcflux} task was used to find the 0.3--10\,keV $3\sigma$ flux upper limits of non-detections, using a 3\arcsec\ radius source region surrounded by a 20\arcsec\ radius annulus for the background. 

\subsection{Swift}

{\it Swift} monitored NGC~6946 closely during 2017, and we use all observations from that year. X-ray source products were generated using the FTOOLS task {\tt xrtpipeline}. The spectrum for ULX\=/3 was taken using a 45\arcsec\ radius source region and a 70\arcsec\ radius background region located outside of the galaxy, and the spectrum was grouped to 20 counts per bin as for {\it NuSTAR}. $3\sigma$ flux upper limits for non-detections in {\it Swift} observations were determined using the FTOOLS task {\tt sosta}, using a 20\arcsec\ radius source region (to avoid contamination by a nearby source) and a 70\arcsec\ radius background region as for ULX\=/3. Magnitude lower limits for optical/ultraviolet non-detections were determined using the FTOOLS task {\tt uvotsource}.

\subsection{HST}

Optical photometry was performed on five observations of NGC~6946 taken using the WFPC2 and WFC3/IR cameras on board the {\it Hubble Space Telescope} ({\it HST}) during three different epochs before 2017, which cover the region of the galaxy in which ULX\=/4 is contained. We use observations in the F547M, F606W and F814W optical bands with WFPC2 (proposal IDs 8591, 8597 and 8599 respectively; the F656N band is also observed, but the resolution is insufficiently good to characterise individual sources in this band) and the F110W and F128N near-infrared bands with WFC3/IR (proposal ID 14156). We retrieved pre-processed images from the {\it Hubble} Legacy Archive created from multiple exposures combined using the MultiDrizzle routine. 

We corrected the {\it HST} astrometry using the USNO star catalogue and the Image Reduction and Analysis Facility (IRAF) tools {\tt ccfind}, {\tt ccmap} and {\tt ccupdatewcs}, and combined the 90\% confidence error circle of the resulting {\it HST} position errors in quadrature with the {\it XMM-Newton} source position error, resulting in a 0.9\arcsec\ 90\% error circle around the source position which we used to identify potential counterparts. We performed aperture and PSF-fitting photometry on these potential optical counterparts using the DAOPHOT-II/ALLSTAR software package \citep{stetson87}, using the values from aperture photometry where a good PSF fit could not be found, and placed magnitude limits based on a combination of the read noise, dark current, and sky background where a source could not be detected at all.

\section{Results and Discussion} \label{sec:results}

All three previously identified ULXs in NGC~6946 are detected by {\it XMM-Newton}, with ULX\=/3 also detected by {\it NuSTAR}. ULX\=/1 and ULX\=/2 were both at sub-ULX luminosities during the simultaneous {\it XMM-Newton} and {\it NuSTAR} observation, and only detected with {\it XMM-Newton}, although we briefly characterise them for completeness. Both {\it XMM-Newton} and {\it NuSTAR} also detect a new ULX, which we call ULX\=/4, at 20:34:56.7 +60:08:12. In this section we present our analysis and discussion of these four objects. We perform all spectral fitting using v12.10 of the XSPEC \citep{arnaud96} software, and all quoted models are given in XSPEC syntax. Uncertainties are given at the 90\% confidence level, and we use the abundance tables of \citet{wilms00} throughout.

\subsection{NGC 6946 ULX-1 and ULX-2}

\begin{figure}
	\begin{center}
	\includegraphics[height=7.5cm]{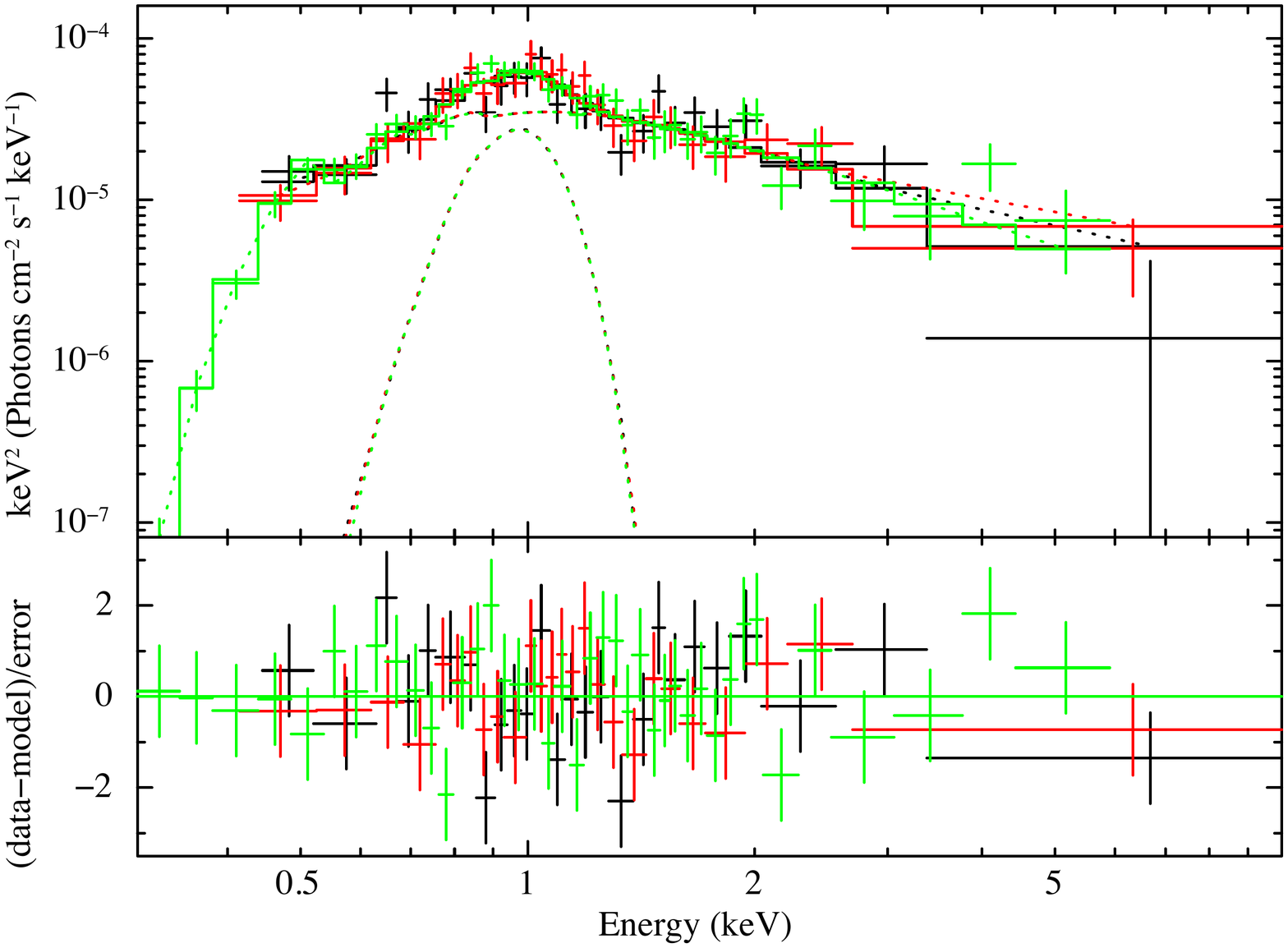} \par
	\vspace{-1cm}
	\includegraphics[height=7.5cm]{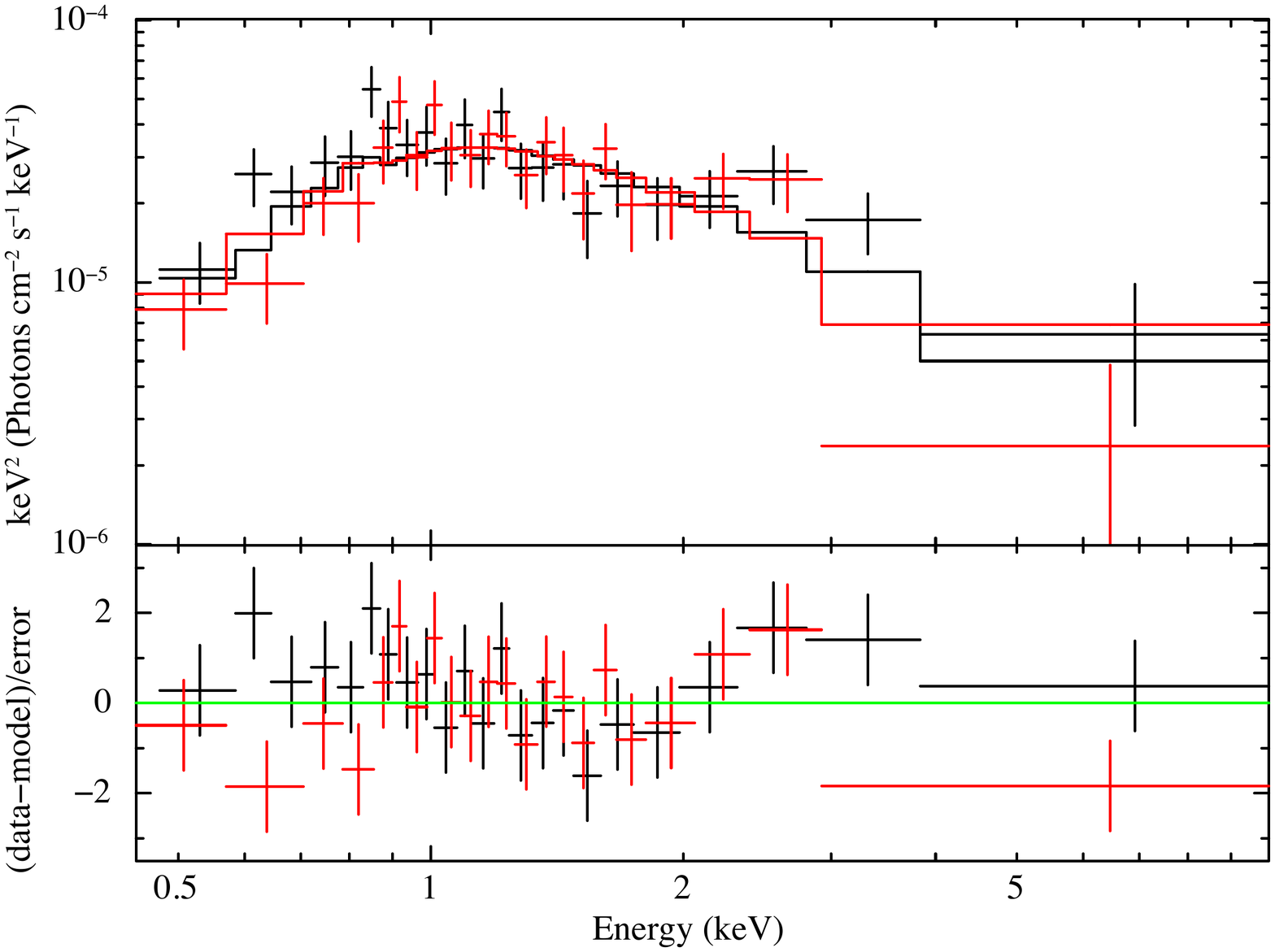}
	\vspace{-1cm}
	\end{center}
	\caption{The {\it XMM-Newton} observation 0794581201 spectrum and residuals for ULX\=/1 ({\it top}) and ULX\=/2 ({\it bottom}). Black, red and green data points are EPIC MOS1, MOS2 and pn data respectively -- ULX\=/2 lies on a pn chip gap, so only MOS data is available. ULX\=/1 is plotted with the best-fitting absorbed power-law and Gaussian model with $\Gamma = 3.6^{+0.4}_{-0.3}$ and $E_{\rm line}=0.90^{+0.04}_{-0.07}$\,keV. ULX\=/2 is plotted with the best-fitting absorbed power-law model with $\Gamma=3.6^{+0.4}_{-0.3}$. Full fit details are provided in Table~\ref{tab:ulx12}. \label{fig:ulx12}}
\end{figure}

We fitted the {\it XMM-Newton} spectra of ULX\=/1 and ULX\=/2 from observation 0794581201 with an absorbed power-law model, using two {\tt tbabs} absorption components, one frozen to the Galactic value of $N_{\rm H} = 1.84\times10^{21}$\,cm$^{-2}$ and the other allowed to vary. In the case of ULX\=/1, a power-law model was not sufficient to produce a good fit, with $\chi^2$/dof = 131.9/90 and the fit showing significant soft residuals at $\sim1$\,keV. These soft residuals are known to be a common feature in the spectra of ULXs and bright X-ray binaries (e.g. \citealt{bauer04,carpano07,middleton15b}), and found in other ULXs to be a combination of emission and absorption features related to powerful outflowing winds (including in NGC~6946~ULX\=/3; \citealt{pinto16}). We used an additional Gaussian component to empirically fit these soft residuals, which resulted in a very large statistical improvement and provided an acceptable fit ($\chi^2$/dof = 84.5/87). Using the best-fitting model for each source, we calculated the absorbed 0.3--10\,keV flux of both objects. We do not correct for absorption since extending steep power-laws, which are empirical rather than physical models, to low energies is likely to overestimate the true flux of the object beneath the absorption -- unabsorbed fluxes using such models can be a factor of two or three higher than those for more physically-motivated models, and the amount of absorption is itself model-dependent, so we make fewer assumptions by just considering the observed, absorbed flux. For comparison, we also fitted both spectra with an absorbed multicolour disc model. Although this was able to produce a statistically acceptable fit, the Gaussian representing the soft residuals in ULX\=/1 contributes to a physically unreasonable portion of the spectrum, broadening to fit the majority of the soft emission, and the residuals for ULX\=/2 are even less well characterised than for a power-law fit. Therefore, a steep power-law seems to be the best empirical model for both spectra. We present the spectral fit results for these two sources in Table~\ref{tab:ulx12} and the best-fitting model plots in Fig.~\ref{fig:ulx12}. 

Both sources are found to be very soft and at a low enough flux to be a little below the ULX threshold luminosity (both have $L_{\rm X}\sim6-7\times10^{38}$\,erg\,s$^{-1}$), consistent with these sources being persistently bright Eddington threshold objects that only occasionally reach the ULX luminosity regime. ULX\=/1 is consistent with a steep power-law spectral shape with soft residuals, as previously found in \citet{earnshaw17}, where it was suggested that it is a super-Eddington accreting source viewed at high accretion rate and/or at a high inclination, with most of the high-energy central emission obscured and down-scattered by the surrounding wind and outer disc. While the most typical examples of ultraluminous supersoft sources have spectra that are almost entirely dominated by thermal emission (e.g. \citealt{urquhart16}), the very steep power-law tail may come from a very small amount of visible central emission, making ULX\=/1 a source on the borderline between the soft ultraluminous regime and the thermal-dominated ultraluminous supersoft sources. ULX\=/2 lies on an EPIC-pn chip gap, so we only have data from the EPIC-MOS cameras. Its spectrum, while similarly steep to ULX\=/1 and statistically well-fitted by a simple absorbed power-law, appears on visual inspection to have a two-component shape to its residuals not dissimilar to that found in the soft ultraluminous accretion regime in ULXs (e.g. \citealt{gladstone09}). However, likely owing to the low signal-to-noise and limited bandpass, fitting the spectrum with two thermal components rather than a power-law does not provide a significant improvement in the quality of fit. Still, it is possible that this source is also a super-Eddington accreting system like ULX\=/1, only with typical luminosity below the standard ULX definition.

\subsection{NGC 6946 ULX-3}

\begin{figure}
	\begin{center}
	\includegraphics[width=7.5cm]{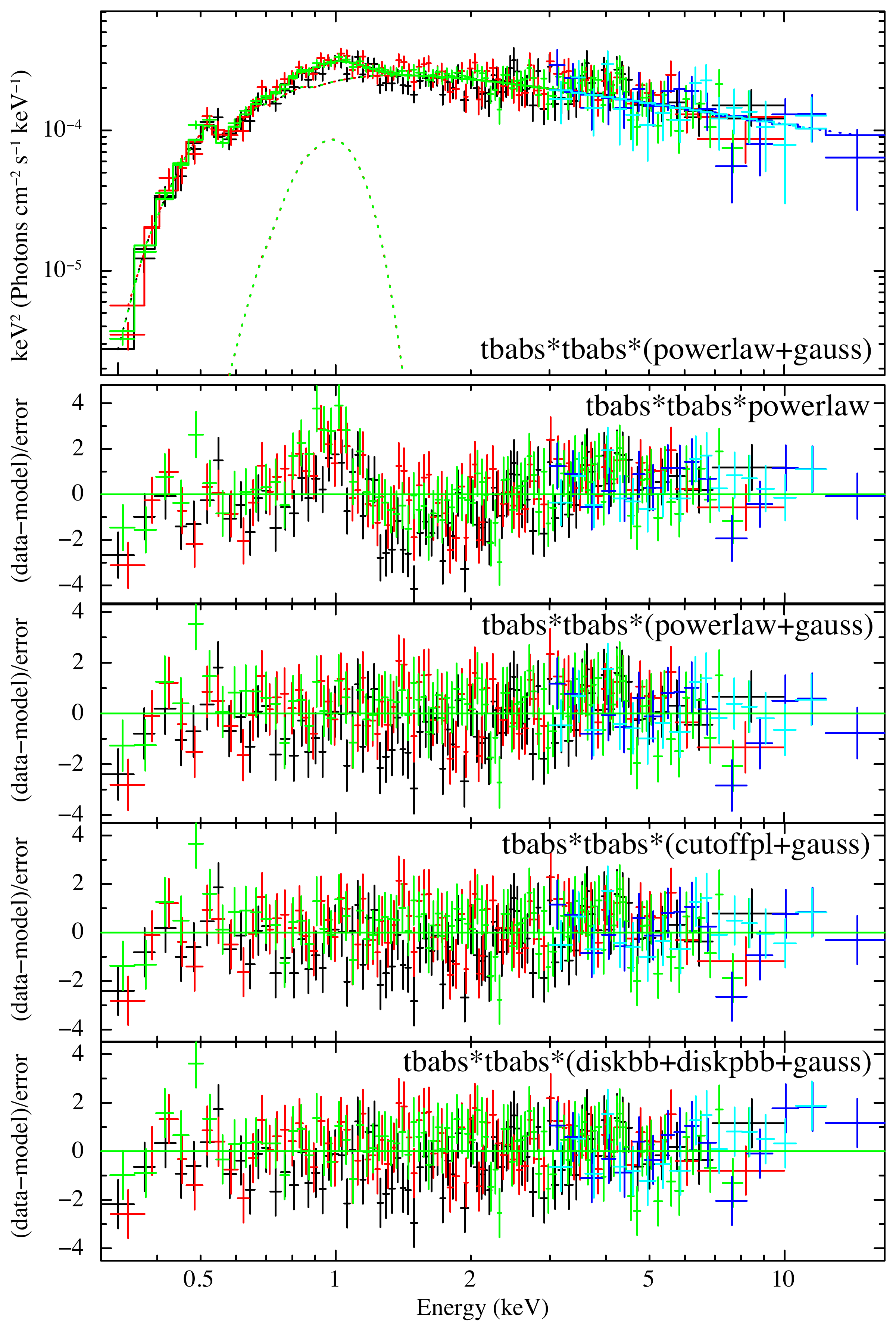}
	\end{center}
	\vspace{-4mm}
	\caption{The {\it XMM-Newton} observation 0794581201 and {\it NuSTAR} observation 90302004004 joint spectrum of ULX\=/3, with {\it XMM-Newton} colours as in Fig.~\ref{fig:ulx12} and the {\it NuSTAR} FPMA and FPMB data plotted in blue and cyan respectively. This spectrum is plotted with the best-fitting absorbed power-law and Gaussian model, with $\Gamma=2.51\pm0.05$ and $E_{\rm line}=0.90^{+0.04}_{-0.06}$\,keV, along with the residuals for four models used to fit the spectrum as labelled in the figure. \label{fig:ulx3}}
\end{figure}

We began by fitting the combined {\it XMM-Newton} and {\it NuSTAR} spectrum of ULX\=/3, with the cross-calibration fixed to unity, with an absorbed power-law model as for ULX\=/1 and ULX\=/2. Like ULX\=/1, its residuals are dominated by the likely wind-related feature at $\sim$1\,keV, so we fitted this feature with a Gaussian component as before, which yields a statistically acceptable fit to the spectrum ($\chi^2$/dof = 332.8/289). With a photon index of $\Gamma=2.51\pm0.05$, ULX\=/3 is the softest ULX with a broadband X-ray spectrum detected by {\it NuSTAR} studied so far. We test for a spectral turnover at high energies as seen in other ULXs, a key indicator of super-Eddington accretion. However, we find that there is no significant improvement in fit with a cut-off power-law model over a power-law without a cut-off ($\Delta\chi^2 = 1.2$ for 1 dof). We simulated 1000 spectra with the cut-off power-law parameters we measure, and found that for a spectrum this soft, {\it NuSTAR} is unable to significantly detect a cut-off and correctly constrain its parameters. We show the best-fitting model parameters for the different models in Table~\ref{tab:ulx3} and show the spectrum and residuals in Fig.~\ref{fig:ulx3}. 

We used the best-fitting power-law model to calculate the absorbed 0.3--10\,keV flux of ULX\=/3 during this observation, which we find to be $(9.1\pm0.2)\times10^{-13}$\,erg\,cm$^{-2}$\,s$^{-1}$, corresponding to a luminosity of $(6.5\pm0.1)\times10^{39}$\,erg\,s$^{-1}$. This is broadly consistent with previously calculated luminosities for this source when accounting for lower distances to the galaxy used previously (e.g. \citealt{middleton15a}), and it therefore continues to be a persistent ULX. 

\begin{deluxetable*}{cccccccccc}
	\vspace{-1mm}
	\tablecaption{The joint ULX\=/3 spectral fitting results for {\it XMM-Newton} observation 0794581201 and {\it NuSTAR} observation 90302004004, and the results for {\it Swift} observation 00010130008 and {\it NuSTAR} observation 90302004002. \label{tab:ulx3}}
	\tablecolumns{10}
	\tablenum{4}
	\tablewidth{0pt}
	\tablehead{
		\colhead{$N_{\rm H}$\tablenotemark{a}} & \colhead{$\Gamma$} & \colhead{$E_{\rm cutoff}$} & \colhead{$T_{\rm in,bb}$} & \colhead{$T_{\rm in,pbb}$} & \colhead{$p$\tablenotemark{b}} & \colhead{$E_{\rm line}$} & \colhead{$\sigma_{\rm line}$} & \colhead{$\chi^2/{\rm dof}$} & \colhead{$F_{0.3-10 \rm keV}$\tablenotemark{c}} \\
		\colhead{($10^{21}$\,cm$^{-2}$)} & \colhead{} & \colhead{(keV)} & \colhead{(keV)} & \colhead{(keV)} & \colhead{} & \colhead{(keV)} & \colhead{(keV)} & \colhead{} & \colhead{($10^{-13}$\,erg\,s$^{-1}$)}
	}
	\startdata
	\multicolumn{10}{c}{{\it XMM-Newton + NuSTAR Epoch 2}} \\
	{\tt tbabs*tbabs*} & \multicolumn{2}{c}{\tt (powerlaw} & & & & \multicolumn{2}{c}{\tt + gauss)} & & \\
	$1.2\pm0.1$ & $2.71\pm0.04$ & - & - & - & - & - & - & 489.8/292 & $8.7\pm0.1$ \\ 
     $0.8\pm0.1$ & $2.51\pm0.05$ & - & - & - & - & $0.90^{+0.04}_{-0.06}$ & $0.16^{+0.04}_{-0.03}$ & 332.8/289 & $9.0\pm0.1$ \\ 
     {\tt tbabs*tbabs*} & \multicolumn{2}{c}{\tt (cutoffpl} & & & & \multicolumn{2}{c}{\tt + gauss)} & & \\
     $0.6\pm0.3$ & $2.4^{+0.1}_{-0.2}$ & $>9.95$ & - & - & -  & $0.85^{+0.06}_{-0.11}$ & $0.20^{+0.07}_{-0.04}$ & 331.6/288 & - \\ 
     {\tt tbabs*tbabs*} & & & {\tt (diskbb} & \multicolumn{2}{c}{\tt + diskpbb} & \multicolumn{2}{c}{\tt + gauss)} & & \\
     $0.7\pm0.3$ & - & - & $0.24^{+0.03}_{-0.02}$ & $2.1\pm0.2$ & $<0.52$ & $0.97^{+0.03}_{-0.04}$ & $0.10\pm0.01$ & 324.8/286 & - \\
     \multicolumn{10}{c}{{\it Swift + NuSTAR Epoch 1}} \\
     {\tt tbabs*tbabs*} & \multicolumn{2}{c}{\tt powerlaw} & & & & & & & \\
     $2\pm2$ & $2.8\pm0.2$ & - & - & - & - & - & - & 48.6/49 & $8.5\pm0.8$ \\
     {\tt tbabs*tbabs*} & \multicolumn{2}{c}{\tt cutoffpl} & & & & & & &  \\
     $<2.8$ & $2.2^{+0.6}_{-0.3}$ & $8^{+63}_{-3}$ & - & - & - & - & - & 45.2/48 & - \\
     {\tt tbabs*tbabs*} & & & {\tt (diskbb} & \multicolumn{2}{c}{\tt + diskpbb)} & & & &  \\
     $<6.2$ & - & - & $0.4^{+0.1}_{-0.2}$ & $2.1^{+0.6}_{-0.7}$ & $<0.9$ & - & - &  43.1/46 & - \\
	\enddata
	\tablenotetext{a}{The column density for the absorption component allowed to vary, as in Table~\ref{tab:ulx12}.}
	\tablenotetext{b}{The radial dependence of disc temperature, as in Table~\ref{tab:ulx12}.}
	\tablenotetext{c}{The absorbed flux in the 0.3--10\,keV band.}
	\vspace{-5mm}
\end{deluxetable*}

\begin{figure}
	\begin{center}
	\includegraphics[width=8cm]{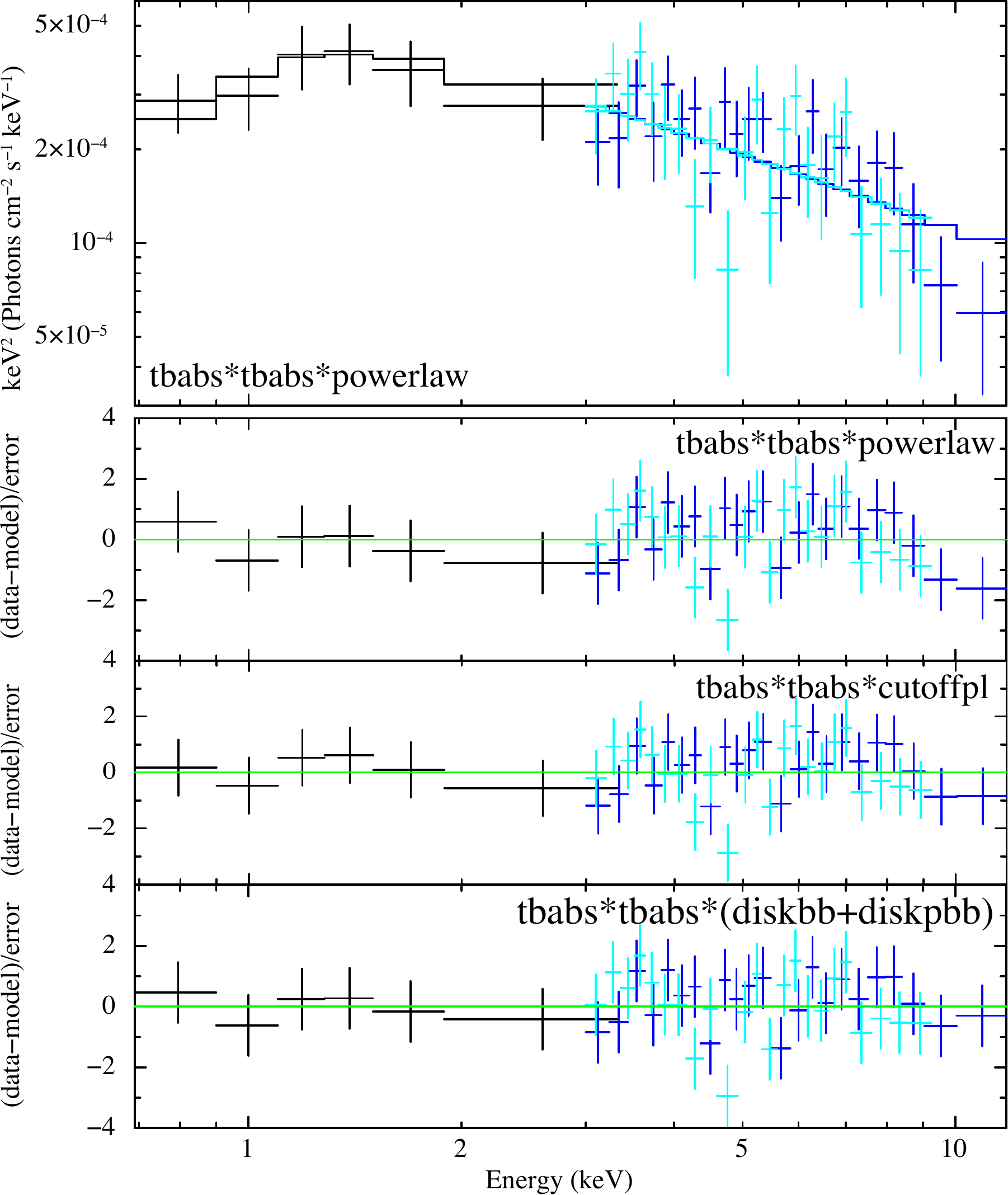}
	\end{center}
	\vspace{-4mm}
	\caption{The {\it Swift} observation 00010130008 and {\it NuSTAR} observation 90302004002 joint spectrum of ULX\=/3, with {\it Swift} XRT data in black and the {\it NuSTAR} data as in Fig.~\ref{fig:ulx3}. The spectrum is plotted with the best-fitting absorbed power-law model, with $\Gamma=2.8\pm0.2$, along with the residuals for three models used to fit the spectrum as labelled in the figure. \label{fig:ulx3swift}}
\end{figure}

While there is no statistical case for a more complex model, we do consider the scenario that this ULX has a similar spectrum to other ULXs observed with {\it NuSTAR} in the past and fit it with a typical {\tt diskbb+diskpbb} model -- the soft {\tt diskbb} component originating in an outer thin disc/soft outflowing wind, and the harder {\tt diskpbb} component originating in the geometrically thick inner disc \citep{walton14,middleton15a}. We find that fitting the spectrum in this manner yields an acceptable fit ($\chi^2$/dof = 324.8/286), but also hints at a hard excess as often seen in hard ULXs at energies $>20$\,keV \citep{mukherjee15,walton14,walton15b}, which cannot be entirely accounted for by a maximally broadened disc component at high energies. This hard excess feature is not significant in these observations of ULX\=/3, but if it is the case that this source is spectrally similar to other ULXs with an excess present, this may be an additional reason why we do not see evidence of a turnover in the spectrum, as the entire spectrum is sufficiently soft that the strength of a turnover feature would be counteracted by a steep power-law excess at the energies we observe. 

There is no simultaneous {\it XMM-Newton} observation for the first {\it NuSTAR} observation, though there is a {\it Swift} observation the following day, so we fit these observations together to get our best possible picture of the broadband spectrum during the first {\it NuSTAR} epoch. The amount of {\it Swift} data is insufficient to determine whether the soft residuals that we fitted with a Gaussian in the {\it XMM-Newton} data are contributing to the soft emission in the first observation as well, so the best comparison we can make is to the power-law-only fit to the second observation. We find that both the flux and spectral hardness are consistent with the second epoch, with no evidence for any significant variability on a ten-day timescale. While the {\it NuSTAR} data appears to hint at a spectral turnover at $\sim8$\,keV on a visual inspection of the residuals, a cut-off power-law offers only a small improvement in $\chi^2$ over a regular power-law model ($\Delta\chi^2 = 3.4$ for 1 dof). We show the best-fitting model parameters in Table~\ref{tab:ulx3} and show the spectrum and residuals in Fig.~\ref{fig:ulx3swift}. 

\begin{figure}
	\vspace{-8mm}
	\begin{center}
	\includegraphics[width=9cm]{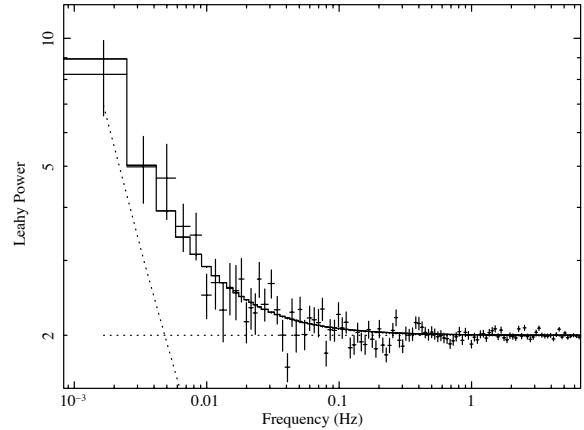}
	\end{center}
	\vspace{-8mm}
	\caption{The EPIC-pn power spectrum for ULX-3 during observation 0794581201, plotted using Leahy normalisation and fitted with a constant for the white noise level and a power-law model for the red noise.  \label{fig:powspec}}
\end{figure}

Our weaker constraints on the soft part of the spectrum in the first epoch limit our ability to comment on any variability of ULX\=/3 between the two observations. However, we can investigate the short-term variability of ULX\=/3 by creating an EPIC-pn power spectrum for the {\it XMM-Newton} observation, averaging the periodograms of 70 segments of length 601\,s (i.e. 8192 time bins at the 73.4\,ms time resolution of the EPIC-pn instrument). We find the power spectrum to exhibit red noise, as expected to be found in accreting systems (see Fig.~\ref{fig:powspec}), but find no evidence for any QPOs at $\sim10^{-2}$\,Hz like those reported for earlier observations in \citet{rao10}. We can rule out the presence of a 8.5\,mHz QPO, as previously observed, at the 5$\sigma$ level for this observation. We also do not significantly detect any other QPOs in the power spectrum. However, the fractional rms variability of ULX-3 is $40\pm4\%$ over the 1--100\,mHz range, showing that ULX-3 continues to be highly variable and its spectrum remains bright and steep as it has previously been observed, indicating that the QPOs are possibly a transient feature in an otherwise persistent accretion state. 

\subsection{NGC 6946 ULX-4}

ULX-4 is a new transient source. It was not detected in the first {\it NuSTAR} observation, but was strongly detected by both telescopes in the second joint {\it XMM-Newton} and {\it NuSTAR} observation.

\begin{figure}[!t]
	\begin{center}
	\includegraphics[width=7.5cm]{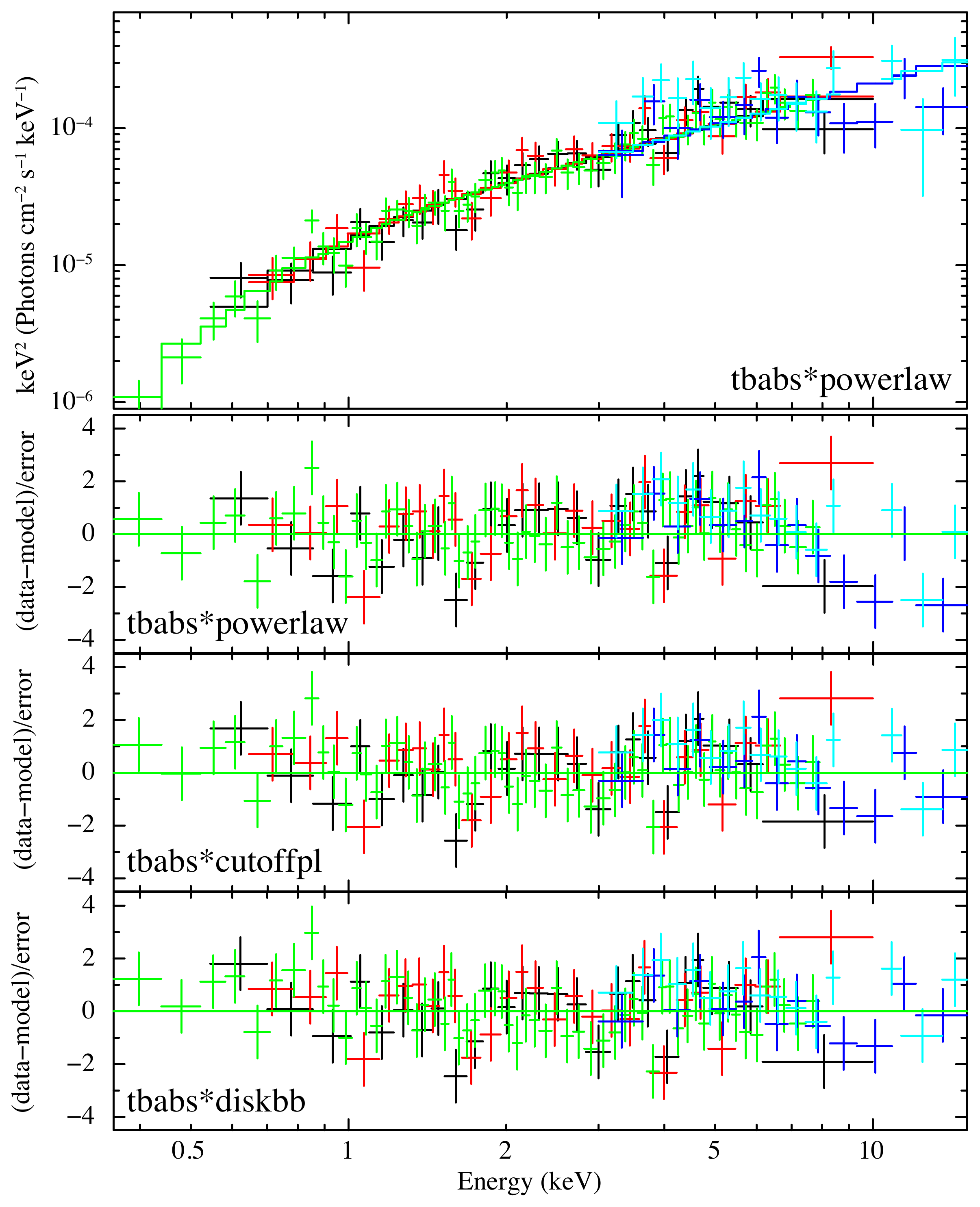}
	\end{center}
	\caption{The {\it XMM-Newton} observation 0794581201 and {\it NuSTAR} observation 90302004004 joint spectrum of ULX\=/4, plotted with the best-fitting absorbed power-law model with $N_{\rm H}$ frozen to the Galactic value and $\Gamma=1.02\pm0.07$, along with residuals for a power-law and a cut-off power-law model. \label{fig:ulx4}}
\end{figure}

As for the other sources, we fitted the combined {\it XMM-Newton} and {\it NuSTAR} spectrum of ULX\=/4 with an absorbed power-law model. We found that no absorption component is required in addition to the Galactic column, so our models for this source only contain a single {\tt tbabs} component frozen to the Galactic value, which is somewhat unusual for ULXs which usually show evidence of a local absorption column. This source has a very hard ($\Gamma\sim1$) power-law-shaped spectrum, and the fit is greatly improved ($\Delta\chi^2=15.8/1$\,dof) by the presence of a cut-off at $\sim11$\,keV. We checked the significance of this cut-off by simulating a power-law spectrum 10,000 times and finding the improvement in $\chi^2$ of a cut-off power-law model compared to a power-law due to random fluctuations in the spectrum. We found that 0.02\% of simulations had $\Delta\chi^2 \geq 15.8$, making the cut-off that we detect in the source spectrum significant to $\sim3.5\sigma$. The spectrum can also be well-fitted with a very hot multicolour disk blackbody model with $kT_{\rm in} = 4.3^{+0.6}_{-0.5}$\,keV, with no need for broadening using a {\tt diskpbb} model -- although over the portion of the spectrum we see, this is functionally very similar to the cut-off power-law model. We find the time-averaged 0.3--10\,keV flux to be $3.2\times10^{-13}$\,erg\,cm$^{-2}$\,s$^{-1}$, placing this new source just inside the ULX regime at $L_{\rm X} = (2.27\pm0.07)\times10^{39}$\,erg\,s$^{-1}$. We show the joint {\it XMM-Newton} and {\it NuSTAR} spectrum in Fig.~\ref{fig:ulx4}, and the spectral fitting results for ULX\=/4 in Table~\ref{tab:ulx4}.

\begin{deluxetable}{ccccc}
	\tablecaption{The ULX\=/4 spectral fitting results for {\it XMM-Newton} observation 0794581201 and {\it NuSTAR} observation 90302004004. \label{tab:ulx4}}
	\tablecolumns{5}
	\tablenum{5}
	\tablewidth{0pt}
	\tablehead{
		 \colhead{$\Gamma$/$T_{\rm in}$} & \colhead{$E_{\rm cut}$} & \colhead{$F_{0.3-10\rm keV}$} & \colhead{$F_{3-20\rm keV}$} & \colhead{$\chi^2/{\rm dof}$} \\
		 \colhead{(-/keV)} & \colhead{(keV)} & \multicolumn{2}{c}{($10^{-13}$\,erg\,cm$^{-2}$\,s$^{-1}$)} & \colhead{} 
	}
	\startdata
	\multicolumn{2}{c}{\tt tbabs*powerlaw} & & & \\
	$1.00\pm0.05$ & - & $3.2\pm0.1$ & $3.9\pm0.3$ & 174.2/143 \\
     \multicolumn{2}{c}{\tt tbabs*cutoffpl} & & & \\
     $0.7\pm0.1$ & $11^{+9}_{-4}$ & $3.2\pm0.1$ & $3.5\pm0.2$ & 158.4/142 \\
     \multicolumn{2}{c}{\tt tbabs*diskbb} & & & \\
     $4.3^{+0.6}_{-0.5}$ & - & $3.0\pm0.1$ & $3.2\pm0.1$ & 158.7/142 \\
	\enddata
\end{deluxetable}

\begin{figure*}[!t]
	\begin{center}
	\includegraphics[height=7cm]{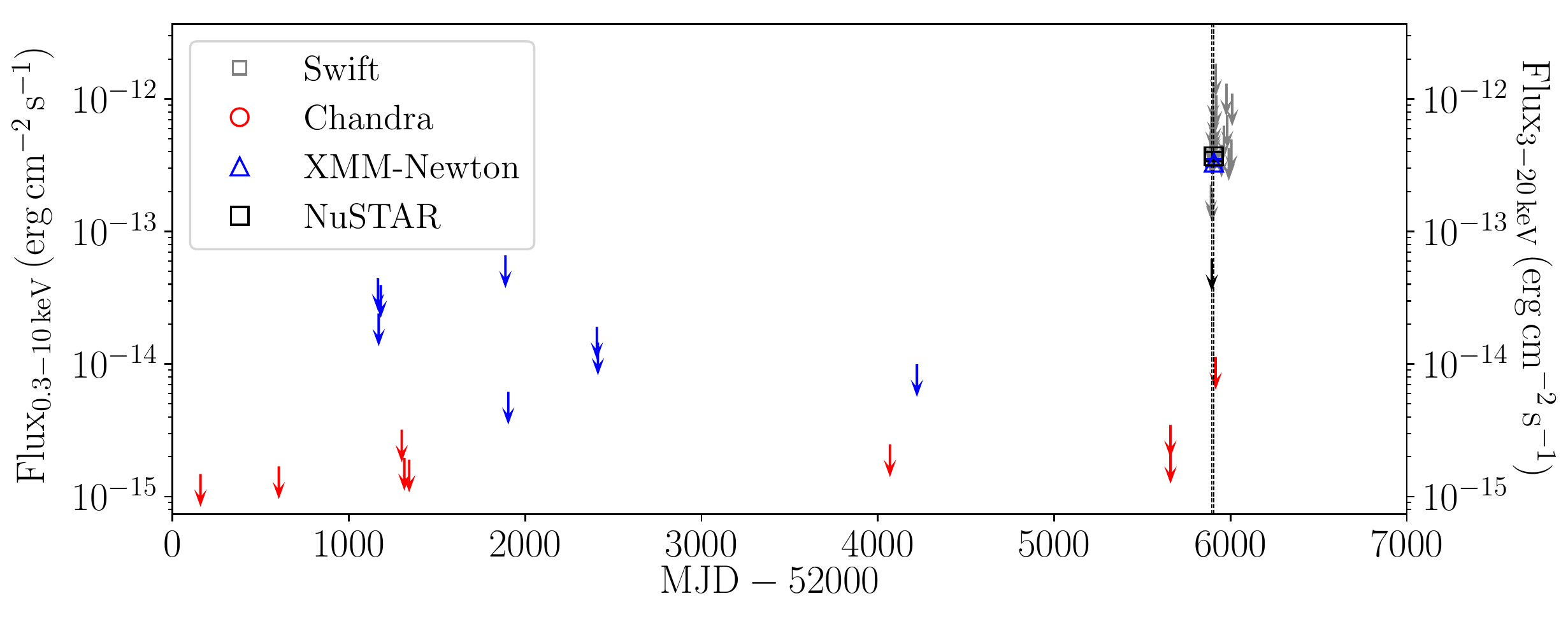}\\
	\includegraphics[height=7cm]{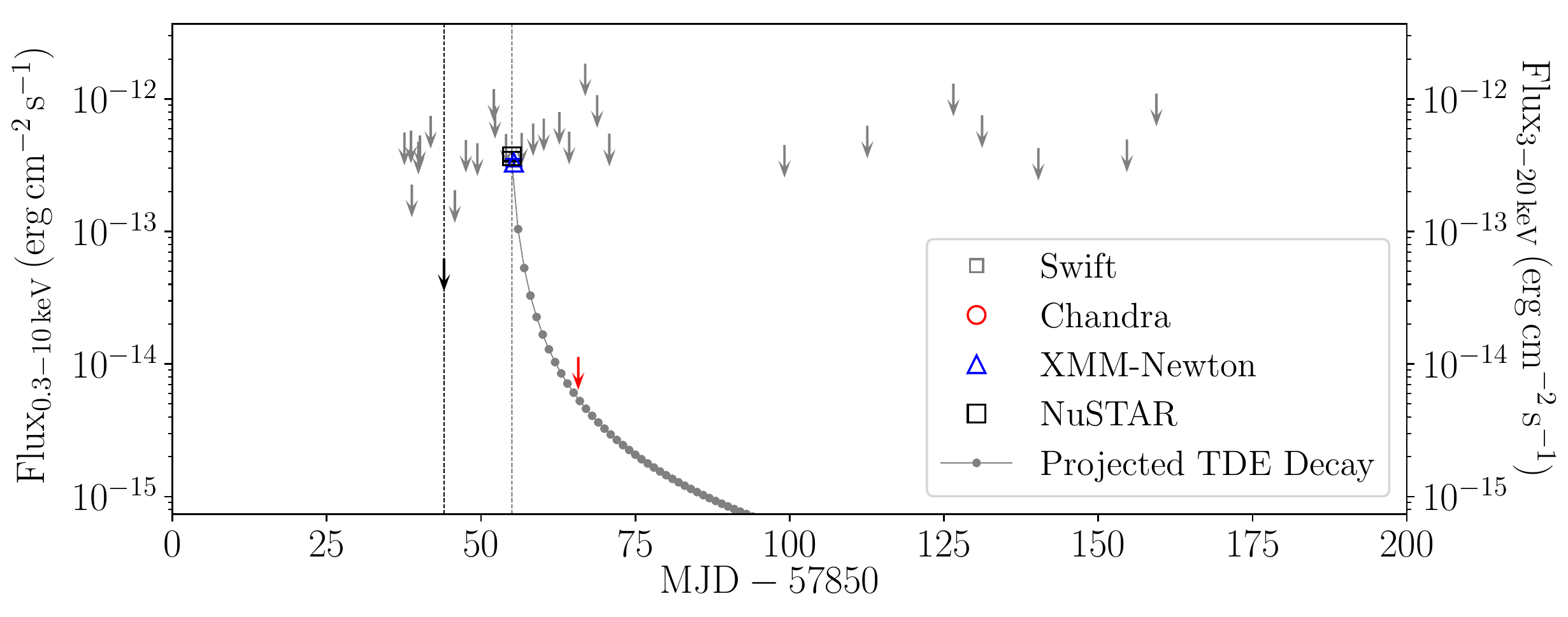}
	\end{center}
	\vspace{-4mm}
	\caption{Long-term lightcurves for ULX\=/4, for the past $\sim17$ years ({\it top}) and for the observations in 2017 ({\it bottom}). The left axis denotes soft (0.3--10\,keV) X-ray flux and applies to {\it XMM-Newton} (blue), {\it Chandra} (red) and {\it Swift} (gray) data points. The right axis denotes hard (3--20\,keV) X-ray flux and applies to {\it NuSTAR} (black) data points. Detections are marked with symbols and upper limits by downward-pointing arrows. The times of the two {\it NuSTAR} observations are indicated by vertical black dashed lines. A $t^{-5/3}$ decay curve typical for tidal disruption events is projected from the {\it XMM-Newton} flux in grey circles.  \label{fig:ulx4lc}}
\end{figure*}

We examined {\it Swift}-XRT data both immediately before, between and after the two {\it NuSTAR} observations, as well as into the remainder of 2017. Due to the short observations and the hardness of the source spectrum, ULX\=/4 is not detected in any of these observations and we are unable to place strong limits on the 0.3--10\,keV flux during the period over which the source appears beyond that it undergoes at least a factor $>1.4$ increase in flux between the two {\it NuSTAR} epochs. Similarly, putting an upper limit on the 3--20\,keV flux from the initial {\it NuSTAR} non-detection ($F_{\rm 3-20 keV} < 6.23\times10^{-14}$\,erg\,cm$^{-2}$\,s$^{-1}$) gives us a factor $>5$ increase in flux for that energy band.

In order to confirm that the detection by {\it XMM-Newton} and {\it NuSTAR} is the first appearance of ULX\=/4, we plot a long-term light curve from archival {\it XMM-Newton} and {\it Chandra} observations (Fig.~\ref{fig:ulx4lc}), assuming a similar spectral shape to the one we observe here of $\Gamma=1$ and Galactic absorption. This assumption gives conservative flux upper limits; should the source have previously been in a softer state, these upper limits would be lower. We find no previous detections of ULX\=/4 -- {\it Chandra} observations place the strongest limits on the 0.3--10\,keV flux, with each upper limit below $F_{\rm X} < 3.5\times10^{-15}$\,erg\,cm$^{-2}$\,s$^{-1}$, or $L_{\rm X} < 2.5\times10^{37}$\,erg\,s$^{-1}$. This implies a factor $\gtrsim80$ increase in flux between its detection in 2017 and when the source was previously observed by {\it Chandra} in 2016. Ten days after the second {\it NuSTAR} epoch, a further observation with {\it Chandra} does not detect the source either, implying an upper limit on the 0.3--10\,keV flux of $F_{\rm X} < 1.1\times10^{-14}$\,erg\,cm$^{-2}$\,s$^{-1}$, or $L_{\rm X} < 7.8\times10^{37}$\,erg\,s$^{-1}$, a factor of $\sim26$ below the flux of the {\it XMM-Newton} detection. Therefore this source is transient, and the outburst that we observe lasts a maximum of 20 days.

\begin{figure}[!t]
	\begin{center}
	\includegraphics[width=9cm]{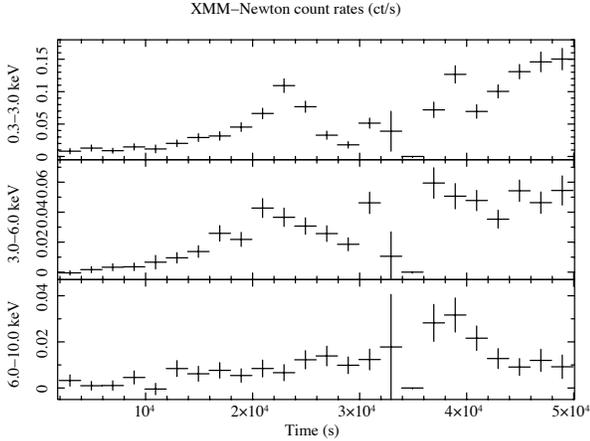} \par
	\end{center}
	\vspace{-10mm}
	\caption{The {\it XMM-Newton} observation 0794581201 light curves for ULX\=/4 in the energy bands 0.3--3.0, 3.0--6.0 and 6.0--10.0\,keV energy bands, in 2000\,s time bins. The drop-out at $3.4\times10^4$\,s corresponds to a period of soft proton flaring which was removed from the analysis and is not due to source behaviour. \label{fig:ulx4xmmlc}}
\end{figure}

\begin{figure}
	\begin{center}
	\includegraphics[width=9cm]{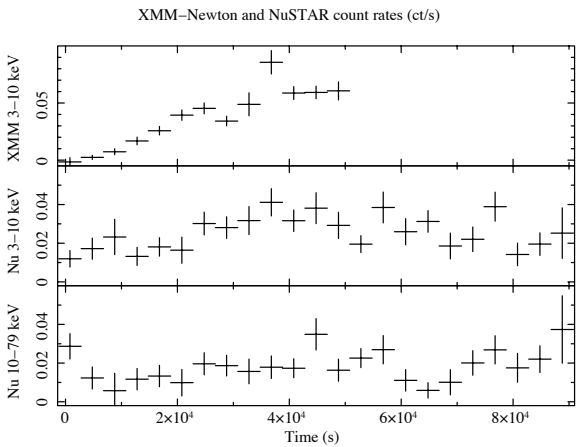} \par
	\end{center}
	\vspace{-10mm}
	\caption{The {\it XMM-Newton} observation 0794581201 light curve for ULX\=/4 in the 3--10\,keV energy band, along with the {\it NuSTAR} light curves in the 3--10\,keV and 10--79\,keV energy bands, in 4000\,s time bins. \label{fig:ulx4xmmnulc}}
\end{figure}

ULX\=/4 is highly variable in the {\it XMM-Newton} energy band over the course of the observation. Its light curve, which we examine in three energy bands (Fig.~\ref{fig:ulx4xmmlc}), shows a low initial flux, a predominantly soft flare lasting $\sim5$\,ks and releasing $\sim$10$^{43}$\,erg of energy, followed by a second increase of flux in the final 10\,ks of the observation with a corresponding softening of the spectrum (there is a gap at $\sim3.4\times10^4$\,s due to removing a period of soft proton flaring rather than source behaviour). The final count rate is over an order of magnitude higher than the count rate at the start of the observation in the soft bands. The {\it NuSTAR} light curve (Fig.~\ref{fig:ulx4xmmnulc}) shows the source to persist throughout the remaining 50\,ks of the observation, though it continues to vary in both soft and hard energy bands.

\begin{figure}[!t]
	\begin{center}
	\includegraphics[width=9cm]{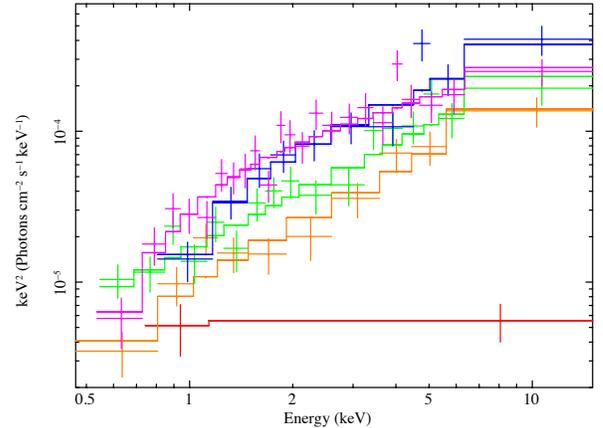}
	\end{center}
	\vspace{-10mm}
	\caption{Time-resolved spectrum of ULX\=/4 from {\it XMM-Newton} observation 0794581201, divided into five 10\,ks intervals, in chronological order of red, orange, green, blue and magenta. For clarity, only EPIC-pn data is shown. \label{fig:timeresspec}}
\end{figure}

Dividing the 50\,ks {\it XMM-Newton} observation into five 10\,ks segments, we can track the spectral and flux evolution of ULX\=/4 over the course of the observation (see Fig.~\ref{fig:timeresspec}). The source begins in a relatively soft low-flux state ($\Gamma=2\pm1$, $F_{\rm X} = 2.6\times10^{-14}$\,erg\,cm$^{-2}$\,s$^{-1}$ in the first 10\,ks of the observation) and quickly reaches its very hard ($\Gamma=0.9\pm0.3$) power-law slope in the second interval. During the third interval, which encompasses most of the soft flare, the spectrum nonetheless remains hard ($\Gamma=0.8^{+0.2}_{-0.1}$) and only shows signs of softening in the final interval ($\Gamma=1.2\pm0.2$). At the end of the observation, the ULX\=/4 has flux $F_{\rm X} = 5.0\times10^{-13}$\,erg\,cm$^{-2}$\,s$^{-1}$. Therefore this source undergoes a factor $\sim20$ increase in flux and transitions into the ULX luminosity regime over the course of the observation.

We created a power spectrum for the {\it XMM-Newton} observation of ULX\=/4 by averaging the periodograms of 72 segments of length 601\,s. There are no significant features in the power spectrum beyond some possible red noise becoming apparent at low frequencies ($\lesssim10^{-3}$\,Hz), which is consistent with an accreting source. Similarly, we created a power spectrum for the {\it NuSTAR} observation, though no features are seen. We searched for pulsations using the High-ENergy Data Reduction Interface from the Command Shell (HENDRICS) software \citep{bachetti15}, a package based on stingray \citep{huppenkothen19} for timing analysis of X-ray data and particularly optimised to handle {\it NuSTAR} data. We ran a pulsation search, considering potential acceleration with $\dot{f}$ up to $10^{-8}$\,Hz\,s$^{-1}$, but failed to detect any pulsations between $10^{-4}$ and 10\,Hz. On simulating light curves of ULX\=/4, assuming a sinusoidal pulsation with a constant 1\,s period at different pulse fractions, we find that we can place a 90\% upper limit on the pulsed fraction of $\sim$20\% for the observed light curve. Given that ULX pulsars are known to be affected by spin-up and orbital modulations (e.g. \citealt{bachetti14,carpano18}), it is conceivable that we would not detect a pulsation at higher pulse fractions than this with the data we currently possess. We therefore cannot rule out the potential presence of pulsations in this source during its outburst.

There is no obvious single optical counterpart to the X-ray source. Data taken with the Ultraviolet and Optical Imager (UVOT) instrument on the {\it Swift} satellite show no detection during any of the {\it Swift} observations of NGC~6946, including the days immediately before and after the second {\it NuSTAR} epoch. While we have no optical data precisely simultaneous with the X-ray detection, this limits the presence of an optical counterpart with magnitude $m_{\rm v}<18.1$ and $m_{\rm b}<19.0$ ($M_{\rm v}<-11.1$, $M_{\rm b}<-10.2$) to less than a day, if one appears at all. However optical counterparts of ULXs are mostly far fainter than this (e.g. \citealt{gladstone13}), so these limits do not place strong constraints on the sort of counterpart that may be present, though they do rule out a very bright optical transient.

\begin{figure*}[!t]
	\begin{center}
	\vspace{-4mm}
	\includegraphics[height=7cm]{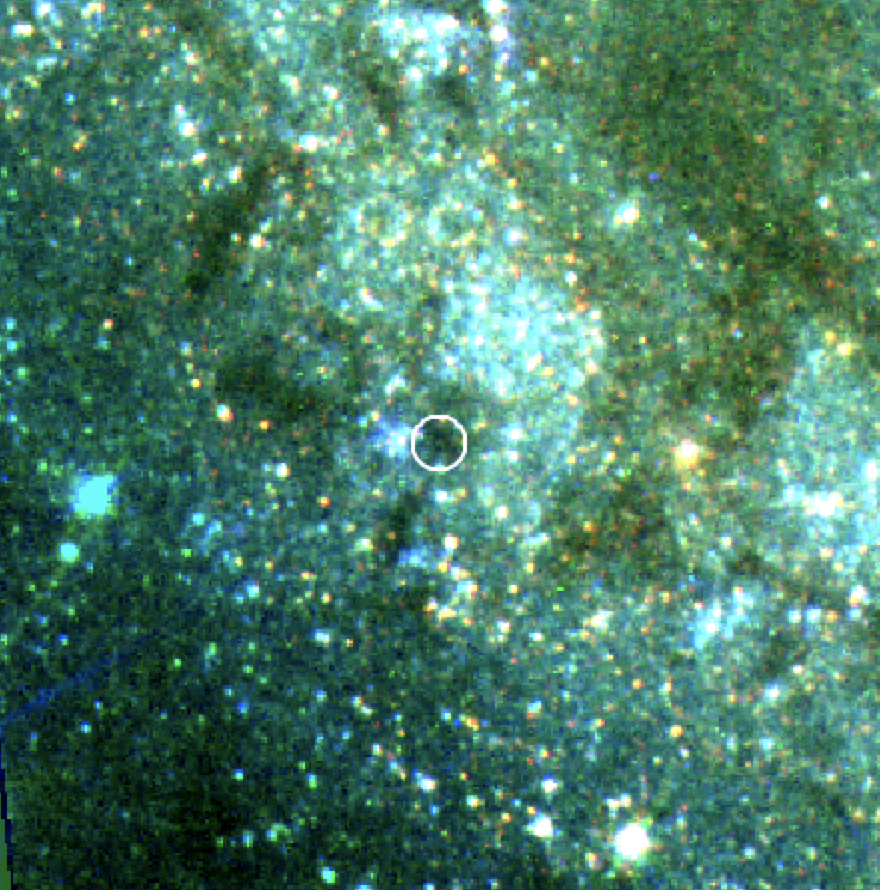}
	\hspace{2mm}
	\includegraphics[height=7cm]{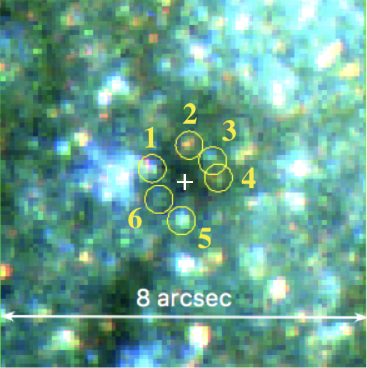}
	\end{center}
	\caption{False-colour {\it HST} images of the region around NGC~6946~ULX\=/4, with the near-infrared F110W band in red, F814W band in green and F606W band in blue. {\it Left}, a $30\times30$\,arcmin square around the source position, marked with a white 0.9\arcsec\ 90\% position error circle. {\it Left}, a closer zoom in on the source region, with six potential optical/infrared counterparts labelled as in Table~\ref{tab:hst}. We note that the IR data is not simultaneous with the optical data. \label{fig:hst}}
\end{figure*}

The region of NGC~6946 in which ULX\=/4 is located has been observed with the Hubble Space Telescope ({\it HST}) on several occasions with the WFPC2 and WFC3/IR instruments -- in the F547M, F606W and F814W optical bands with WFPC2 and the F110W and F128N near-infrared bands with WFC3/IR. The F656N band is also observed with WFPC2, but the spatial resolution is insufficient for performing photometry on the individual sources. All of these observations took place before 2017, but we can still investigate potential optical/infrared counterparts before the outburst. We corrected the astrometry (see Section~\ref{sec:data}) and derived a 0.9\arcsec\ error circle around the {\it XMM-Newton} source position, which we found to be centred on a dark dust cloud. We find six potential counterparts around the edge of this error circle and a bright star cluster just outside it. We show the potential counterparts in Fig.~\ref{fig:hst} in three of the {\it HST} bands and list them in Table~\ref{tab:hst} -- not all sources are visible in all bands. It is also feasible that the source is associated with the star cluster itself, although it is too crowded for us to characterise individual stars within it.

We used the DAOPHOT-II/ALLSTAR software \citep{stetson87} to obtain photometric data for all six sources in the bands that they are detected in, and correct for Galactic reddening using $E(B-V)=0.2942\pm0.0028$ \citep{schlafly11} and the \citet{fitzpatrick99} reddening law. These sources are faint and in a crowded field, and given the presence of a dust lane, may be affected by significant local extinction. Therefore they are not detected at high significance, and these magnitudes should be considered approximations with large errors, especially at the lowest fluxes. We present a table of magnitudes in Table~\ref{tab:hst}. Using a distance modulus of $\mu=29.16$ we convert these to absolute magnitudes. In addition, we use the F606W and F814W bands to produce a $V-I$ colour. 


\subsubsection{Foreground/background source}

ULX\=/4, if definitely located in its apparent host galaxy, is a highly unusual and interesting source. However. we must first assess whether it is a foreground or background contaminant.

Given NGC~6946's position relatively close to the Galactic plane, there is a higher density of foreground sources in its direction compared with galaxies at higher Galactic latitudes, therefore this is a possibility worth investigating. In addition, ULX\=/4 does not exhibit any significant absorption in addition to the Galactic column. This is unusual for ULXs, which tend to have local column densities $\sim10^{21}$\,cm$^{-2}$ due to local absorption within their host galaxies \citep{winter07}. 

\begin{deluxetable*}{cccccccccc}
	\tablecaption{The position and optical/IR magntitudes of potential counterparts to ULX\=/4 observed with {\it HST} within a 0.9\arcsec\ error circle of the {\it XMM-Newton} source position. \label{tab:hst}}
	\tablecolumns{10}
	\tablenum{6}
	\tablewidth{0pt}
	\tablehead{
		 \colhead{ID} & \colhead{Position (J2000)} & \colhead{$m_{\rm 547M}$} & \colhead{$m_{\rm 606W}$} & \colhead{$m_{\rm 814W}$} & \colhead{$m_{\rm 110W}$} & \colhead{$m_{\rm 128N}$} & \colhead{$M_{606}$} & \colhead{$M_{\rm 814}$} & \colhead{$V-I$} \\
		 \colhead{} & \colhead{} &  \colhead{(2001-04-22)} & \multicolumn{2}{c}{(2000-12-25)} & \multicolumn{2}{c}{(2016-02-09)} & &
	}
	\startdata
	1 & 20:34:56.99 +60:08:13.37 & $24.8\pm0.3$ & $23.0\pm0.6$ & $24.1\pm0.3$ & $20.6\pm0.2$ & $23.7\pm0.3$ & $-6.2\pm0.6$ & $-5.0\pm0.3$ & $-1.2\pm0.7$ \\
	2 & 20:34:56.88 +60:08:13.88 & $>26.4$ & $25\pm2$ & $26\pm2$ & $20.9\pm0.1$ & $23.8\pm0.3$ & $-4\pm2$ & $-3\pm2$ & $-1\pm3$ \\
	3 & 20:34:56.82 +60:08:13.54 & $24.9\pm0.5$ & $24.2\pm0.8$ & $25\pm1$ & $24\pm4$ & $26.3\pm1$ & $-4.9\pm0.8$ & $-4\pm1$ & $-1\pm1$ \\
	4 & 20:34:56.80 +60:08:13.17 & $>26.4$ & $>26.7$ & $26\pm4$ & $22.0\pm0.1$ & $25.1\pm0.1$ & $>-2.4$ & $-3\pm4$ & $>0.6$ \\ 
	5 & 20:34:56.90 +60:08:12.25 & $24.0\pm0.2$ & $23.0\pm0.5$ & $23.3\pm0.4$ & $21.5\pm0.1$ & $24.8\pm0.1$ & $-6.2\pm0.5$ & $-5.9\pm0.4$ & $-0.3\pm0.6$ \\ 
	6 & 20:34:56.97 +60:08:12.71 & $26\pm1$ & $24.3\pm0.8$ & $26\pm2$ & $>28.3$ & $25.2\pm0.2$ & $-4.8\pm0.8$ & $-3\pm2$ & $-2\pm2$ \\ 
	\enddata
	\tablecomments{Magnitudes and estimated standard errors are obtained from DAOPHOT II, or lower limits where the source was unable to be characterised with DAOPHOT II. Absolute magnitudes are calculated assuming a distance modulus $\mu=29.16$. The date that each {\it HST} observation was taken is given in the header.}
\end{deluxetable*}

If ULX\=/4 is a foreground source located within the Milky Way, it is very faint -- at a distance of 10\,kpc, its luminosity would be $3.4\times10^{33}$\,erg\,s$^{-1}$. Some Galactic sources such as V404~Cyg and Aql~X\=/1 have quiescent luminosities similar to this (e.g. \citealt{garcia01,campana14}), but since ULX\=/4 has not previously been detected, implying a lower luminosity state with $L_{\rm X}\lesssim10^{31}$\,erg\,s$^{-1}$ during the deepest of these observations, ULX\=/4 must be undergoing some form of outburst rather than being detected in a quiescent state. However, its detected luminosity is low for an outburst, even for very faint X-ray transients which tend to have luminosities in the range of $10^{34}$--$10^{36}$\,erg\,s$^{-1}$ (e.g. \citealt{degenaar12}) during an outburst. Conversely, it is too high a luminosity for most cataclysmic variables aside from intermediate polars (IPs; e.g. \citealt{sazonov06}). IPs can possess hard spectra like that we find for ULX\=/4, but they are also expected to exhibit iron features (e.g. \citealt{kuulkers06,middleton12}) which we do not see in this source. In addition, the lack of an obvious foreground star detection limits any possible Galactic optical counterpart to dwarf stars. While we cannot entirely rule out ULX\=/4 being a foreground source on this basis, it would be an unusual source if it were.

ULX-4's coincidence with a dense, dusty region of one of NGC~6946's spiral arms, and lack of X-ray absorption beyond Galactic levels, suggests that it is likely not a background source either. Its outburst timescale is too short for a standard tidal disruption event \citep{komossa15} and too long for an X-ray flash or fast X-ray transient resulting from an off-axis gamma-ray burst (e.g. \citealt{heise01,yamazaki02}), and too hard a spectrum for either case. 

Therefore we can be reasonably confident that ULX\=/4 is located within NGC~6946 rather than being a foreground or background source coincident with the galaxy.

\subsubsection{Supernova}

With its sudden appearance and increase in X-ray luminosity, we must also consider whether this source is an explosive transient. X-ray emission from core-collapse supernovae can have a range of luminosities, and the luminosity we observe for ULX\=/4 is a reasonable value for a supernova. However, the speed and extent of its subsequent decline in luminosity is far faster than expected for most supernova (e.g. \citealt{dwarkadas12}), and there is no optical counterpart to the X-ray transient, nor a known optical transient at this position in the days before it appeared. Additionally, the short-timescale variability we observe supports the interpretation of ULX\=/4 being an accretion-powered object. Therefore we consider accretion-related explanations for this source.


\subsubsection{Super-Eddington accreting source}

It has become generally accepted that the majority of ULXs are stellar-mass compact objects accreting in a super-Eddington regime, and since ULX\=/4 reaches luminosities above $10^{39}$\,erg\,s$^{-1}$, this is a natural option to consider. Its hard spectrum, with evidence of a turnover at $\sim$10\,keV, shows similarities with other ULXs in a hard ultraluminous regime (e.g. \citealt{sutton13a,walton18a}), some of which are confirmed NS ULXs. 


If ULX\=/4 is such a source, this serendipitous detection offers unprecedented insight into the onset of super-Eddington accretion, as the source moves from a sub-Eddington to a super-Eddington state over the course of the observation. We find that, aside from a brief soft flare at early times, the onset of accretion is entirely dominated by hard emission, with no evidence of a significant disc component -- though the signal quality may not be sufficient to detect the presence of a relatively dim disc. The spectrum as a whole only begins to show signs of softening at the end of the {\it XMM-Newton} observation, when the source is firmly in the ULX luminosity regime. This might suggest that the formation of a hot inner disc or accretion column precedes the build-up of an outer disc in this case -- and with the timescale of the event being so short, it is possible that there was insufficient material available to form any kind of large disc before all matter was accreted. This is in contrast with the broadened disc state seen in much of the lower-luminosity ULX population and suggested to be objects accreting at $\sim$Eddington rates, possibly in the process of transitioning into a super-Eddington ultraluminous state (e.g. \citealt{sutton13b}). 

One other factor to consider is the transient nature of this source, being active for only $\sim$10 days, assuming from the rise in luminosity over the observation that we are observing the source as it first appears. While many are highly variable, the majority of ULXs are reasonably persistent sources. One exception to this is the class of confirmed NS ULXs, most of which demonstrate periods of dramatic decrease in flux of well over an order of magnitude (e.g. \citealt{walton15a,fuerst16,israel17b}), potentially due to the source entering the propeller regime. With the small sample size of sources demonstrating such behaviour, the duty cycle of accretion and propeller state in these sources is not well understood. It is feasible that we are instead observing a NS briefly leaving and then returning to a propeller state, allowing only a brief onset of accretion. Since there are no previous observations of this source, it may be the case that we have witnessed its first onset of accretion, having been initially produced with a sufficiently high magnetic field and/or rotation frequency to begin in the propeller regime. However, our previous coverage of the source is not sufficient to rule out previous short outbursts in the past if the source simply has a low duty cycle.

All the potential optical counterparts we have identified in the {\it HST} data are consistent in brightness and colour with being bright blue main sequence or OB-supergiant stars -- as they are potentially affected by local extinction due to the presence of a dust lane, the latter case is more likely -- except for source 4, and potentially 1 and 2, which have properties more consistent with red supergiants. Bright optical counterparts with $-8 < M_V < -4$ are not uncommon for ULXs, though these counterparts may have significant contribution to their optical emission from the irradiated accretion disc rather than the companion star (e.g. \citealt{grise12,fabrika15,ambrosi18}). We have no evidence that ULX\=/4 was accreting at the dates of the {\it HST} observations (although none of the observations were simultaneous with an X-ray observation so neither can we definitively rule this out). Under the assumption that ULX\=/4 is a new appearance, a blue accretion disc-dominated optical companion is less likely in this case, but this source may well be a high-mass X-ray binary with an optical counterpart dominated by the companion star. Red supergiants are also viable ULX counterparts (e.g. \citealt{heida14}), so none of the potential counterparts can be ruled out on that basis. It is possible that future {\it HST} observations will reveal changes in the optical/NIR emission that may aid in identifying a single counterpart.

A NS ULX briefly leaving the propeller regime is a reasonable explanation for this transient source. However, there are no pulsations detected from ULX\=/4 and no other definitive evidence that it is a NS. Therefore, we also explore other scenarios that could explain our results.

\subsubsection{Transient outbursts}

Some transient ULXs are thought to be classical X-ray binary outbursts that happen to be luminous enough to briefly reach the ULX regime before declining in flux again (e.g. \citealt{middleton13}). While these sources usually undergo a transition from a low/hard to a high/soft state over the course of the outburst, the recent WATCHDOG survey (\citealt{tetarenko16}, and references therein) suggests that up to 40\% of outbursts in Galactic BH binaries do not reach the soft state, with their spectra remaining hard over the duration of the outburst. Such hard transients do not always follow the fast-rise exponential-decay pattern seen in the more typical `canonical' outbursts (e.g. \citealt{brocksopp04}), and have timescales from the tens to hundreds of days.

However, there are several problems with a fast ``hard-only'' outburst scenario for ULX\=/4. Firstly, the spectrum of ULX\=/4 is unusually hard even for such outbursting sources, which tend to have photon indices typical for the sub-Eddington hard state of $\Gamma=1.4$--1.7 (e.g. \citealt{revnivtsev00,belloni02,sidoli11}). Also, hard-only outbursts do not tend to reach the fluxes that full outbursts do, generally only reaching Eddington fractions of $\sim10\%$ \citep{tetarenko16}. Such an outburst would require ULX\=/4 to be an intermediate-mass BH (IMBH) with mass $\sim200$\,M$_{\odot}$. It is possible for transient BHs in the hard state to reach up to 100\% of Eddington before a transition to the soft thermal state \citep{dunn10}, though there is no evidence for such a transition taking place and it is unlikely that such an outburst could have concluded by the time of the latest {\it Chandra} non-detection of ULX\=/4. Therefore we conclude that this kind of hard-only outburst is unlikely to be the cause of our detection of ULX\=/4.

The known source that may be a potential analogue to ULX\=/4 is V404~Cyg, a low-mass X-ray binary that went into an outburst lasting tens of days in 2015 after being in a quiescent state for 26 years. Instead of exhibiting the sub-Eddington states expected during most X-ray binary outbursts, it appeared to enter a super-Eddington regime, with a highly-variable and often very hard spectrum, especially at its highest fluxes \citep{motta17b}. While the average luminosity over the course of the entire outburst was not super-Eddington, reaching $L_{\rm X}>10^{39}$\,erg\,s$^{-1}$ only during brief peaks in emission, much of its strong variability in flux and spectral shape can be attributed to varying levels of absorption of a slim disc inner accretion flow (e.g. \citealt{sanchezfernandez17}), suggesting that its intrinsic luminosity may have been more consistently high. It does differ from ULX\=/4 in its strong emission in the {\it INTEGRAL} band, with a reflection spectrum up to 100\,keV \citep{motta17a} rather than the lower-energy turnover seen in ULXs. Nevertheless, the precedent for relatively short-duration super-Eddington outbursts accompanied by long periods of quiescence exists, and this may be a plausible explantion for ULX\=/4's behaviour. 


\subsubsection{Micro-tidal disruption event}


We have only observed ULX\=/4 once, so we cannot rule out a one-time transient scenario without a second detection of the source. A tidal disruption event (TDE) is one such scenario, and occurs when a star is disrupted and accreted onto a super-massive BH. Most TDEs have a spectrum that peaks in the ultraviolet or soft X-rays, and decline on a timescale of months to years with a characteristic $t^{-5/3}$ power-law drop-off in flux due to the rate of matter fallback (for a recent review of TDE observations, see \citealt{komossa15}).

As this source is not at the centre of its host galaxy and possesses a hard spectrum rather than a soft one, a typical TDE does not appear to be a good match for our observations of ULX\=/4. However, it has been suggested that micro-tidal disruption events ($\mu$TDEs), in which a low-mass star or large planet is disrupted or partially disrupted by a stellar-mass BH or IMBH, may also occur and may possess different observational signatures than typical TDEs (e.g. \citealt{perets16}). 

If we assume that ULX\=/4 reaches its peak luminosity during the joint {\it XMM-Newton} and {\it NuSTAR} observation and decays according to a typical $t^{-5/3}$ power-law immediately afterwards, we find that the {\it Chandra} non-detection upper limit is actually consistent with such a scenario (see Fig.~\ref{fig:ulx4lc}). Even if there is a delay before the source flux begins to decay, some simulations indicate that a small BH mass or an ultra-close encounter with an IMBH could lead to a high fallback rate, early intersection of disrupted matter and very rapid accretion disc formation in close proximity to the BH (possibly subject to general relativistic effects), instead of the $t^{-5/3}$ decline from typical self-interaction of the disrupted matter (e.g. \citealt{evans15,kawana18}). The rapid accumulation of an accretion disc would lead to a brief period of super-Eddington accretion in which the disc matter is drained, with a timescale of $10^5$--$10^6$\,s (e.g. \citealt{perets16}). This is consistent with our observation of ULX\=/4. 

Some fast and bright X-ray transients have previously been attributed to the potential tidal disruption of a star by an IMBH (e.g. \citealt{jonker13}, though this example is still brighter and softer than ULX\=/4). We can also compare ULX\=/4 to well-studied central TDEs such as the J1644+57 event in 2011 \citep{burrows11,levan11}, which is particularly relevant as a ``jetted TDE'' with properties different from more typical TDEs -- namely a hard spectrum and a super-Eddington luminosity (a second source, J2058+05, has similar properties and is also thought to be a jetted TDE; \citealt{cenko12}). It has been argued that J1644+57 demonstrates the properties of a scaled-up, transient ULX (e.g. \citealt{socrates12,kara16}), so it may be the case that the transient ULX\=/4 is a scaled-down example of a similar event. Its luminosity was far higher than ULX\=/4's, reaching $\sim$10$^{48}$\,erg\,s$^{-1}$ at its brightest, and a naive scaling of mass with peak luminosity would suggest a compact object a fraction of the mass of the Sun for ULX\=/4, so there would have to be a substantial difference in Eddington ratio in play as well if these sources have a similar origin. In early times it demonstrated flaring behaviour on $\sim$hour timescales, to which the flare during ULX\=/4's increase in brightness could be analogous. ULX\=/4's spectrum is far harder even than that of J1644+57, which at $\Gamma\sim2$ is already harder than all other TDEs observed to date \citep{auchettl17}. However, J1644+57 does demonstrate a softening of its spectrum with its drop in flux, exhibiting a photon index closer to $\Gamma\sim1.5$ at its highest fluxes at early times. Therefore it may be feasible for a $\mu$TDE to exhibit a very hard spectrum at its earliest times. Finally, J1644+57 as a jetted TDE is also distinct from other TDEs in that it has a very high X-ray to optical flux ratio, evidence of less reprocessing of X-ray emission into the optical regime. It also possesses the lowest column densities compared to other TDEs. The lack of a bright optical counterpart to ULX\=/4 and the absence of an absorption component beyond the Galactic contribution in ULX\=/4 bears some similarities to this scenario.


Should ULX\=/4 be a super-Eddington $\mu$TDE, the observed luminosity would imply a BH mass of $\sim$10\,M$_{\odot}$, which is low to induce tidal disruption in most cases but may still do so for a low mass object. Under the assumption that the highest luminosity we observe is the peak luminosity of a $\mu$TDE, $L_{\rm peak} = 3.2\times10^{39}$\,erg\,s$^{-1}$, this is consistent with the luminosity expected from the tidal disruption of a $\sim$0.01\,M$_{\odot}$ brown dwarf by a 16\,M$_{\odot}$ BH. 


Assuming a $t^{-5/3}$ decline from a maximum flux observed during the joint {\it XMM-Newton} and {\it NuSTAR} observation and no further cut-off, we project the flux at the time of writing this paper to be $<10^{-17}$\,erg\,s$^{-1}$, so unfortunately the opportunity to feasibly confirm such a decay law has long since passed. However with the data we currently possess, a $\mu$TDE analogous to the super-Eddington jetted TDE scenario, with a stellar-mass BH disrupting a brown dwarf and undergoing a subsequent brief period of super-Eddington accretion, is a plausible explanation for ULX\=/4. 


\section{Conclusions} \label{sec:conc}

NGC~6946 remains a valuable galaxy for the study of super-Eddington accretion. Both ULX\=/1 and ULX\=/2 possess very steep power-law spectra and are potential examples of ultraluminous supersoft sources dominated by outer disc and wind emission, even though their luminosity is usually in the Eddington threshold regime of $10^{38}$--$10^{39}$\,erg\,s$^{-1}$. In fact, NGC~6946 contains other sources within this luminosity regime which may also offer examples of similar behaviour, and further opportunities to study the effects of super-Eddington accretion beyond the ULX population. While their lower luminosity makes them more challenging than ULXs to study, the reasonably low distance to NGC~6946 makes it possible to place decent constraints on the spectrum with moderate-to-long observations with {\it XMM-Newton}, and will make them a highly interesting population to examine with next-generation X-ray missions such as {\it Athena} \citep{nandra13}.

Given that ULX\=/3 was not the primary target, the 2017 {\it NuSTAR} observations were not optimised to characterise it. Even so, despite it being a reasonably soft ULX with $\Gamma=2.51\pm0.05$, {\it NuSTAR} is able to detect it almost to 20\,keV. We do not significantly detect a high-energy turnover as seen in the spectra of other ULXs, but due to the softness of the spectrum and the potential presence of a steep power-law after such a break (as seen in the best quality {\it NuSTAR} data of hard ULXs), identifying such a break would be difficult. While {\it NuSTAR} is most suitable for studying ULXs with hard spectra, this does go to show that softer ULXs can also be detected and studied with this mission. 

The serendipitous discovery of ULX\=/4 provides a fascinating potential example of the onset of super-Eddington accretion. Its very hard spectrum, equally consistent with a cut-off power-law model and a hot disc blackbody, and transient nature make it challenging to identify. However, it bears similarity to the observed hard spectra of neutron star ULXs, which also have the capacity to undergo sudden and dramatic rises and falls in flux due to the propeller effect, although we are unable to detect pulsations from ULX\=/4 (and would not necessarily expect to given the number of photons observed from the source). Additionally, the brief duration of its appearance and lack of previous detections raise questions about whether the object has a particularly low duty cycle or whether it is a new system. Alternatively, the outburst could be consistent with a super-Eddington outburst similar to V404~Cyg, although it does not exhibit the same reflection-dominated spectrum at high energies, or a micro-tidal disruption event in which a low-mass object such as a brown dwarf is disrupted by a stellar-mass compact object, although its spectrum is harder even than the jetted TDEs previously observed. Ultimately, further detections of ULX\=/4 would be required to establish whether it is a NS ULX or other relatively persistently accreting object rather than a transient event.

\acknowledgments

We thank our anonymous referee for useful comments on this paper. This work was supported under NASA contract NNG08FD60C. DJW acknowledges financial support from STFC in the form of an Ernest Rutherford fellowship. This work made use of data from the {\it NuSTAR} mission, a project led by the California Institute of Technology, managed by the Jet Propulsion Laboratory, and funded by the National Aeronautics and Space Administration. This work has also made use of observations by {\it XMM-Newton}, an ESA science mission with instruments and contributions directly funded by ESA Member States and NASA. Results reported in this article are based in part on public data obtained from the {\it Chandra} and {\it Swift} data archives, and on observations made with the NASA/ESA {\it Hubble Space Telescope}, obtained from the Data Archive at the Space Telescope Science Institute, which is operated by the Association of Universities for Research in Astronomy, Inc., under NASA contract NAS 5-26555.

%

\vspace{5mm}
\facilities{NuSTAR, XMM, CXO, Swift(XRT), HST}
\software{astropy \citep{astropy13,astropy18}, CIAO \citep{fruscione06}, HENDRICS \citep{bachetti15}, HEASoft \citep{heasarc14}, NuSTARDAS, {\it XMM-Newton} SAS}


\bibliography{ngc6946ulxpaper}
\bibliographystyle{../aasjournal}

\end{document}